\newif\ifPhIIpreprint
\PhIIpreprinttrue 
\newif\ifArxiv
\Arxivtrue 
\ifPhIIpreprint\else
    \Arxivfalse 
\fi

\ifPhIIpreprint
    \documentclass[%
     reprint,
     amsmath,amssymb,
     aps,
     prx,
    ]{revtex4-2}
\else
    \documentclass[9pt,twocolumn,twoside]{pnas-new}
    \templatetype{pnasresearcharticle} 
\fi

\usepackage{graphicx}
\usepackage{dcolumn}
\usepackage{bm}
\usepackage{mathrsfs}
\usepackage{comment}
\usepackage{mathtools}
\usepackage{afterpage}
\usepackage{booktabs}
\usepackage{xcolor}
\usepackage{soul}
\usepackage{tensor}
\usepackage{amsmath}
\usepackage{amsthm}
\usepackage{hyperref}
\usepackage{enumitem}
\usepackage{tikz}
\usepackage{accents}
\hypersetup{colorlinks,linkcolor={blue!90!black},citecolor={blue!90!black},urlcolor={blue!90!black}}
\usepackage[normalem]{ulem}
\usepackage{float}
\usepackage{makecell}
\usepackage{cleveref}
\ifPhIIpreprint
    \usepackage{pdfpages}
    \makeatletter
    \AtBeginDocument{\let\LS@rot\@undefined}
    \makeatother
    \usepackage{atbegshi}
\fi

\DeclareMathAlphabet\mathbfcal{OMS}{cmsy}{b}{n}

\newcommand{\norm}[1]{\left\lVert#1\right\rVert}
\newcommand{\R}{\mathbb{R}}
\newcommand{\um}{$\mu$m}
\newcommand{\umsq}{$\mu$m$^2$}

\newcommand{\mmcu}{mm$^3$}

\def\skb#1{\textcolor{blue}{\normalsize #1}}

\ifPhIIpreprint
    \newcommand{\SI}{SI Appendix}
    \newcommand{\dropcap}[1]{#1}
    \newcommand{\matmethods}[1]{\section*{Materials and Methods}#1}
    \newcommand{\showmatmethods}{}
    \newcommand{\acknow}[1]{\section*{Acknowledgements}#1}
    \newcommand{\showacknow}{}
    
    
\else
    \newcommand{\SI}{SI Appendix}
\fi

\begin{document}

\title{Spatiotemporal distribution of the glycoprotein pherophorin II reveals stochastic 
geometry of the growing ECM of \textit{Volvox carteri}}

\ifPhIIpreprint
    \author{Benjamin von der Heyde${}^{1}$}
    \thanks{Joint first author}
    \author{Anand Srinivasan${}^{2}$}
    \thanks{Joint first author}
    \author{Sumit Kumar Birwa${}^{2}$}
    \author{Eva Laura von der Heyde${}^{1}$}
    \author{Steph S.M.H. H{\"o}hn${}^{2}$}
    \email[]{sh753@cam.ac.uk}
    \author{Raymond E. Goldstein${}^{2}$}
    \email[]{R.E.Goldstein@damtp.cam.ac.uk}
    \author{Armin Hallmann${}^{1}$} 
    \email[]{armin.hallmann@uni-bielefeld.de}
    
    \affiliation{\vspace{2mm}${}^{1}$Department of Cellular and Developmental Biology of Plants, University of 
    Bielefeld, Universit{\"a}tsstr. 25, 33615 Bielefeld, Germany}
    \affiliation{\vspace{1mm}${}^{2}$Department of Applied Mathematics and Theoretical 
    Physics, Centre for Mathematical Sciences,\\ University of Cambridge, Wilberforce Road, Cambridge CB3 0WA, 
    United Kingdom}
\else
    \author[a,1]{Benjamin von der Heyde}
    \author[b,1]{Anand Srinivasan}
    \author[b]{Sumit Kumar Birwa}
    \author[a]{Eva Laura von der Heyde}
    \author[b,2]{Steph S.M.H. H{\"o}hn}
    \author[b,2]{Raymond E. Goldstein}
    \author[a,2]{Armin Hallmann} 
    \affil[a]{Department of Cellular and Developmental Biology of Plants, University of 
    Bielefeld, Universit{\"a}tsstr. 25, 33615 Bielefeld, Germany}
    \affil[b]{Department of Applied Mathematics and Theoretical 
    Physics, Centre for Mathematical Sciences,\\ University of Cambridge, Wilberforce Road, Cambridge CB3 0WA, 
    United Kingdom}
    \leadauthor{von der Heyde}
\fi

\ifPhIIpreprint\else
    \significancestatement{The extracellular matrix (ECM) plays many 
    important structural, developmental and physiologic roles in animals, fungi, plants and algae, and was particularly important in evolutionary transitions from unicellular 
    to multicellular organisms. As the ECM is by definition external to cells, there 
    must necessarily be an aspect of self-assembly involved in its generation, yet little is 
    known regarding such processes even in the simplest multicellular species.  Here we report 
    the development of a transgenic strain of the multicellular green alga \textit{Volvox carteri}, 
    in which a key ECM protein, pherophorin II, is fused with yellow fluorescent protein, 
    allowing the first quantitative study of stochastic ECM geometry and growth dynamics 
    in this system, which reveals behavior that is reminiscent of yet distinct from the 
    hydration of foams.}
    
    \authorcontributions{All authors designed research, performed research,
    analyzed data and
    wrote the paper.}
    \authordeclaration{The authors declare no competing interests.}
    \equalauthors{\textsuperscript{1} B.v.d.H. and A.S. contributed equally to this work.}
    \correspondingauthor{\textsuperscript{2}Emails: sh753@cam.ac.uk, reg53@cam.ac.uk \& armin.hallmann@uni-bielefeld.de}

    \keywords{Extracellular Matrix $|$ \textit{Volvox carteri} $|$ Geometry}
\fi

\begin{abstract}
\ifPhIIpreprint Abstract: \fi 
The evolution of multicellularity involved the transformation of a simple cell wall of unicellular ancestors 
into a complex, multifunctional extracellular matrix (ECM). A suitable model organism to study the formation 
and expansion of an ECM during ontogenesis is the multicellular green alga \textit{Volvox carteri}, which,
along with the related volvocine algae, produces a complex, self-organized ECM composed 
of multiple substructures.
These self-assembled ECMs primarily consist of hydroxyproline-rich glycoproteins, a major
component of which is pherophorins. To investigate the geometry of the growing ECM, 
we fused the \textit{yfp} gene with the gene for pherophorin II (PhII) 
in \textit{V. carteri}.
Confocal microscopy reveals PhII:YFP localization at key structures within the ECM, 
including the boundaries of compartments surrounding each somatic cell and the outer surface of the organism.
Image analysis during the life cycle allows the stochastic geometry of those growing
compartments to be quantified.  We find that their areas and aspect ratios exhibit robust gamma distributions and exhibit a transition from a tight polygonal to a looser acircular packing geometry with stable eccentricity 
over time, evoking parallels and distinctions with the behavior of hydrated foams.  These results provide a quantitative benchmark for addressing
a general, open question in biology: How do cells produce structures external to themselves in a robust
and accurate manner? 
\end{abstract}

\ifPhIIpreprint
    \date{\today}
\else
    \dates{This manuscript was compiled on \today}
    \doi{\url{www.pnas.org/cgi/doi/10.1073/pnas.XXXXXXXXXX          }}
\fi

\maketitle

\ifPhIIpreprint\else
    \thispagestyle{firststyle}
    \ifthenelse{\boolean{shortarticle}}{\ifthenelse{\boolean{singlecolumn}}{\abscontentformatted}{\abscontent}}{}
    
    \firstpage[5]{3}
\fi

\dropcap{T}hroughout the history of life, one of the most significant evolutionary transitions was the 
formation of multicellular eukaryotes. 
In most lineages that evolved multicellularity, including animals, fungi and plants, the 
extracellular matrix (ECM) has been a key mediator of this transition by connecting, positioning, and 
shielding cells \cite{abedin2010diverse,stavolone2017extracellular,kloareg2021role,domozych2024extracellular}.
The same holds for multicellular algae such as \textit{Volvox carteri} (Chlorophyta) and its multicellular relatives 
within volvocine green algae which developed a remarkable array of advanced traits in a comparatively short amount 
of evolutionary time\textemdash oogamy, asymmetric cell division, germ-soma division of labor, embryonic 
morphogenesis and a complex ECM \cite{RN4413,RN164,RN896,RN3520}\textemdash 
rendering them uniquely suited model systems for examining evolution from a unicellular progenitor 
to multicellular organisms with different cell types
\cite{RN4413,RN164,RN3520,RN4692,RN5666}.
In particular, \textit{V. carteri}'s distinct and multilayered ECM makes it a model organism 
for investigating the mechanisms underlying ECM growth and its effects on the positioning of 
the cells that secrete its components.
Building on recently established protocols for stable expression of 
fluorescently labeled proteins in \textit{V. carteri} \cite{RN5456,RN5656,RN5682} we present here a new transgenic strain revealing localization of the 
glycoprotein pherophorin II and the first \textit{in vivo} study of the stochastic geometry of a growing ECM.

\begin{figure}[t]
\begin{center}
\includegraphics[width=\columnwidth]{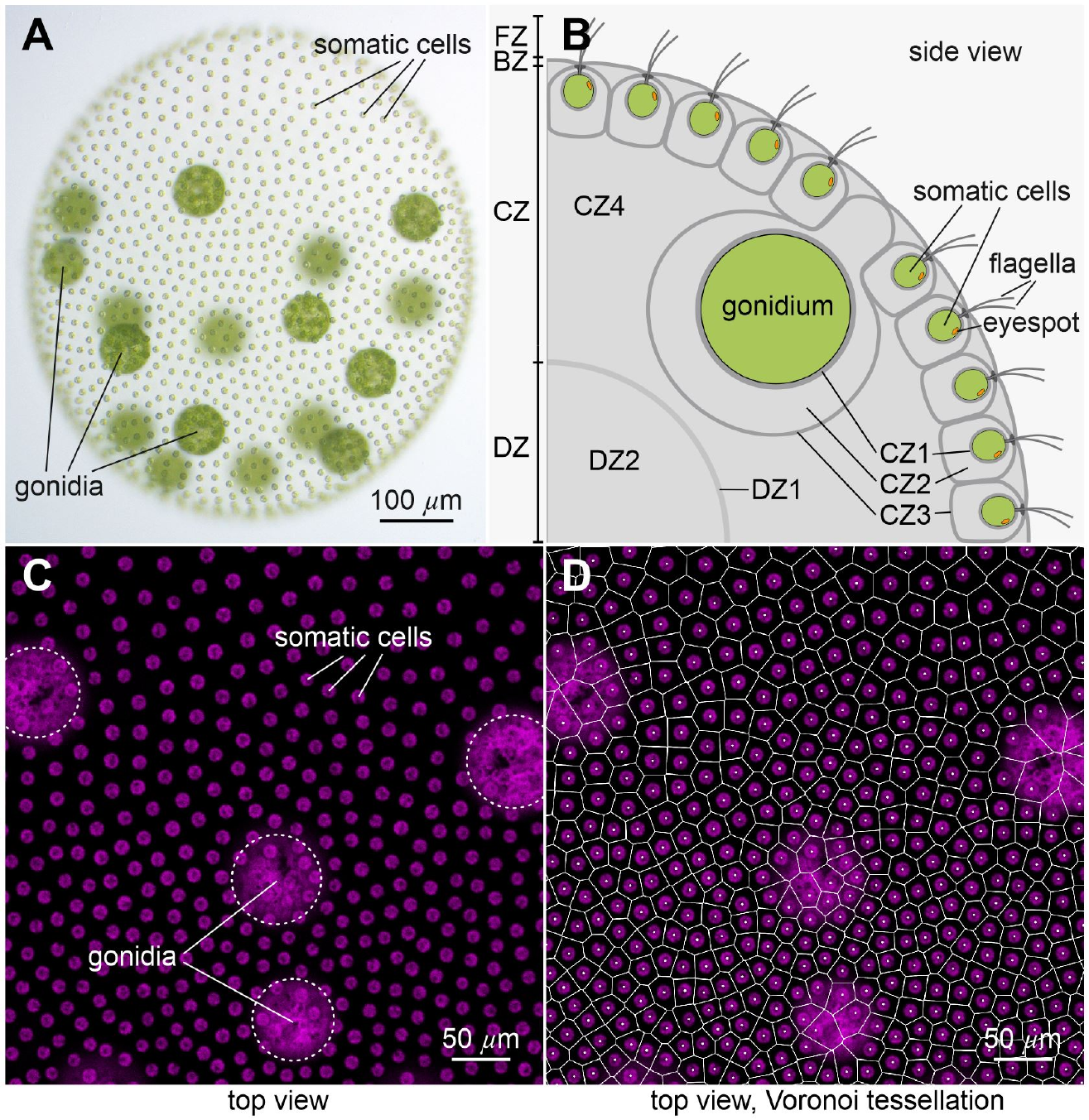}
\caption{Phenotype, ECM architecture, and cell distribution of \textit{V. carteri}.
(A) Adult female (wild-type).  Somatic cells each bear an eyespot and two flagella,
while deeper-lying gonidia are flagella-less. 
All cells have a single large chlorophyll-containing chloroplast (green). 
(B) Zones of the ECM, as described in text. (C) Image utilizing
autofluorescence of the chlorophyll (magenta) to define cell positions. (D) Semi-automated 
readout of somatic 
cell positions in (C) followed by a Voronoi tessellation gives an estimate of 
the ECM neighborhoods of somatic cells.}
\label{fig1}
\end{center}
\end{figure}

\textit{V. carteri} usually reproduces asexually (Fig. \ref{fig1}A).  Sexual development is triggered 
by exposure to heat or a species-specific glycoprotein sex inducer, which results in development of 
sperm-packet-bearing males and egg-bearing females 
\cite{RN616,RN3,RN79,RN5458}. 
In the usual asexual development, a \textit{V. carteri} organism consists of $\sim 2000$ biflagellate somatic cells
resembling \textit{Chlamydomonas} in their morphology, arranged in a monolayer at the surface of a sphere 
and approximately 16 much larger, nonmotile, asexual reproductive cells (gonidia) that constitute the germline, 
lying just below the somatic cell layer \cite{RN4413,RN164,RN566,RN896,RN703}.
The somatic cells are specialized for ECM biosynthesis, photoreception and motility; 
for phototaxis, they
must be 
positioned correctly within the ECM at the surface of the organism \cite{RN616,RN78,RN4}.

The ECM of \textit{V. carteri} has been studied in the past decades from the perspective of structure and 
composition, developmental, mechanical properties, cellular interactions, molecular biology and 
genetics, and evolution \cite{RN616,RN984,RN78,RN4,RN5456}.  
In the ontogenesis of \textit{V. carteri}, ECM biosynthesis in juveniles, inside their mothers, begins 
only after all cell divisions and the process of embryonic inversion are completed. 
Cells of the juveniles then continuously excrete large quantities of ECM building blocks which are integrated  into their ECM, 
producing an enormous increase in the size of the juveniles; within 48 hours the volume increases almost $3000$-fold
as the diameter increases 
from $\sim 70\,\mu$m to $\sim 1\,$mm, raising the question of how relative cell positions are 
affected by growth. 
In the adult organism, the ECM accounts for up to 99$\%$ of the organism's volume and consists of morphologically distinct structures with a defined spatial arrangement.
Based on electron microscopy, a nomenclature was established \cite{RN78}
that defines four main ECM zones: flagellar zone (FZ), boundary zone (BZ), cellular zone (CZ) and 
deep zone (DZ), which are further subdivided (Fig. \ref{fig1}B). The CZ3 forms the ECM compartment boundaries of 
individual cells and the BZ constitutes the outer surface of the organism. CZ3 and BZ show a higher electron 
density than the ECM within the compartments or the ECM in the interior below the cell layer 
(CZ4 and DZ2) \mbox{\cite{RN164,RN896,RN78}}. CZ3 and BZ therefore appear to consist of a firmer, more robust 
ECM, while CZ4 and DZ2 appear to be more gelatinous.  

The ECM of volvocine algae predominantly consists of hydroxyproline-rich glycoproteins (HRGPs), which 
are also a major component of the ECM of embryophytic land plants 
\cite{RN616,RN934,RN5420,RN5421,RN4}, but contains no cellulose.
In \textit{V. carteri} and other volvocine algae, a large family of HRGPs, the pherophorins, are expected to constitute the main building material of the ECM
\mbox{\cite{RN583,RN19,RN9,RN6,RN984,RN5458}.}
Just in \textit{V. carteri}, 118 members of the pherophorin family were discovered in the genome \cite{RN3259,RN5456};
pherophorin genes are typically expressed in a cell type-specific manner
\cite{RN5557} that
can be constitutive or induced either by the sex inducer or wounding
\cite{RN583,RN9,RN6,RN984,RN5458,RN166}.
Pherophorins have a dumbbell-like structure: two globular domains separated by a rod-shaped, 
highly proline-rich domain that varies considerably in length 
\cite{RN9,RN6,RN616,RN4}.
These prolines undergo post-translational modification to hydroxyprolines; pherophorins are also strongly glycosylated
\cite{RN616,RN984,RN4,RN24,RN146}.
For two pherophorins the polymerization into an insoluble fibrous network was shown in vitro 
\cite{RN583}; some have already been localized in the ECM or in ECM fractions 
\cite{RN19,RN6,RN984,RN5456,RN5682}.

To investigate the ECM, we 
determined that pherophorin II (PhII) was the appropriate 
one to label fluorescently, as it is firmly integrated into the ECM.
It is a 70 kDa glycoprotein with strong and immediate inducibility by the sex inducer 
\cite{RN5458,RN163,RN228}.
Because it can only be extracted under harsh conditions, it is thought to be a component of the 
insoluble part of the somatic CZ \cite{RN5458,RN163,RN228}.
We fused one of the nine \cite{RN9,RN3259} gene copies of PhII with the \textit{yfp} gene. The corresponding 
DNA construct was stably integrated into the genome of \textit{V. carteri} transformants by particle bombardment. 
The expression of fluorescent PhII:YFP was confirmed in transgenic mutants through CLSM,
where it was found in the compartment boundaries of the cellular and boundary zones. 
This allows the proportion of ECM secreted by individual somatic cells
to be determined along with the stochastic 
geometry of these ECM structures during development. 

We suggest that a detailed study of the geometry of ECM structures during growth in \textit{Volvox} made possible
by this strain will provide
insight into a more general question in biology: How do cells robustly produce structures 
external to themselves?  There is no physical picture of how the intricate geometry of the
\textit{Volvox} ECM arises through what must be a self-organized process of polymer crosslinking \cite{sumper2000self}.  
While information on a structure's growth dynamics can be inferred 
from its evolving shape, as done for animal epithelial cells \cite{dicko2017geometry},
this connection has not been made for
\textit{Volvox}.

Recently \cite{Day2022}, the somatic cell locations of \textit{V. carteri} were determined 
using their chlorophyll autofluorescence, from which the ECM neighborhoods were determined by 2D Voronoi tessellation, 
as in Figs. \ref{fig1}C,D.
While the somatic cells appear to be arranged in a quasi-regular pattern, this analysis reveals that
their neighborhood areas exhibited a broad, skewed 
k-gamma distribution. Subsequent work has shown that such distributions may arise from 
bursty ECM production at the single-cell level \cite{RN5866}. While the nearest-space-allocation rule assumed by Voronoi tessellations is a reasonable first approximation of the geometry of the somatic CZ3, the validity of this approximation remains unclear.
To investigate this and to answer the more general questions raised above, we performed geometric 
analyses of the structures illuminated by localized PhII:YFP, using metrics (Table \ref{table:metrics}) which give insight into the coexistence of 
local geometric stochasticity and global robustness during growth. In particular, we draw parallels between the temporal evolution 
of ECM structures and wet foams. 
    
\begin{figure}[t]
\begin{center}
\includegraphics[width=\columnwidth]{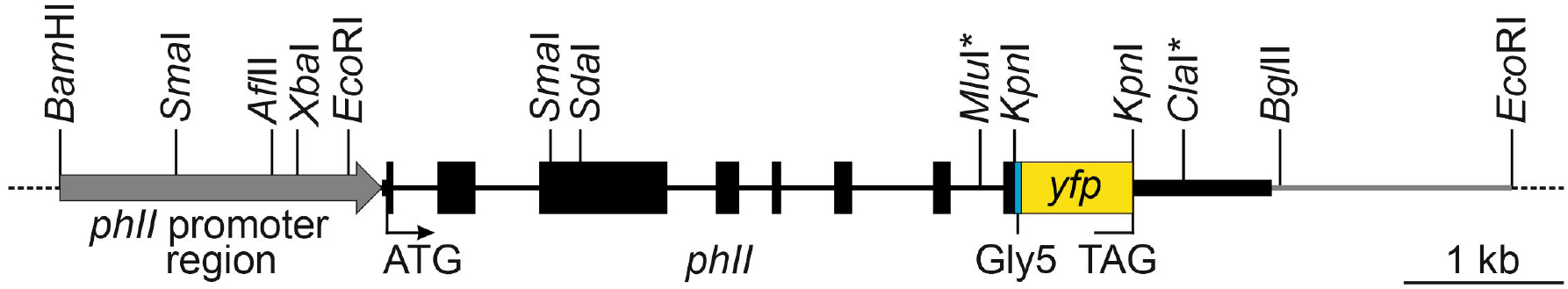}
\caption{Schematic diagram of the transformation vector pPhII-YFP. The vector carries the 
complete \textit{phII} gene including its promoter region. Exons (introns) are shown as black boxes 
(thin lines). Directly upstream of the TAG stop codon, a 0.7 kb fragment coding 
for a flexible penta-glycine 
spacer (cyan) and the codon-adapted coding sequence of \textit{yfp} (mVenus) (yellow) were 
inserted in frame 
using an artificially introduced \textit{Kpn}I site. The \textit{Kpn}I was inserted before 
into the section 
between \textit{Mlu}I and \textit{Cla}I (asterisks). The vector backbone (dashed lines) 
comes from pUC18. For details, see Methods; the sequence of the vector insert is in \SI,
Fig. S3. 
}
\label{fig2}
\end{center}
\end{figure}

\section*{Results}

\subsection{Vector construction and generation of transformants expressing PhII:YFP}  
The pherophorin II gene (\textit{phII}) was cloned from \textit{V. carteri} genomic DNA including its 
promoter, 5' and 3' UTRs and all seven introns (Fig. \ref{fig2}). A sequence coding for a flexible penta-glycine 
spacer and the codon-adapted coding sequence of \textit{yfp} were inserted directly upstream of the \textit{phII} 
stop codon to produce a \textit{phII-yfp} gene fusion. The obtained vector pPhII-YFP (Fig. \ref{fig2}) was 
sequenced and then used for stable nuclear transformation of the nitrate reductase-deficient 
\textit{V. carteri} recipient strain TNit-1013 by particle bombardment. To allow for selection, the 
non-selectable vector pPhII-YFP was co-transformed with the selectable marker vector pVcNR15, which 
carries an intact \textit{V. carteri} nitrate reductase gene and, thus, complements the nitrate reductase 
deficiency of strain TNit-1013. Screening for transformants was then achieved by using medium with nitrate 
as nitrogen source. The  transformants were investigated for stable genomic integration of the 
vector and, via confocal microscopy, for expression of the fluorescent protein at sufficient levels throughout their life cycle. 

\subsection{\textit{In vivo} localization of pherophorin II}   

\begin{figure}[t]
\begin{center}
\includegraphics[width=\columnwidth]{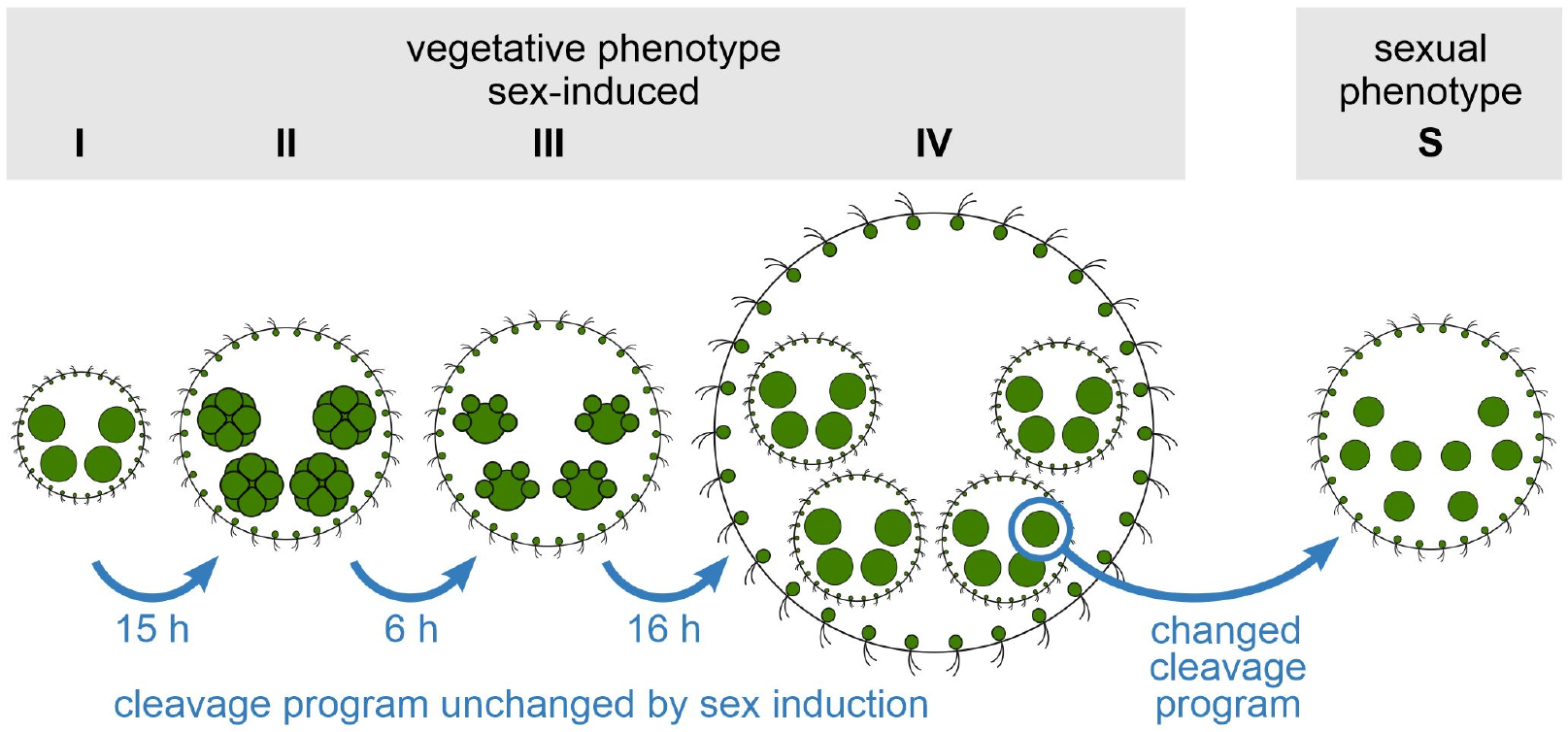}
\caption{Stages of development of \textit{V. carteri} females. 
Localization of PhII and quantification of ECM features were studied in $5$ stages, four with vegetative (asexual) phenotype (I-IV) and one with 
the fully developed sexual phenotype (S). In stages I-IV, incubation with the sex inducer was short enough to increase PhII expression for adequate localization without changing the vegetative cleavage program.}
\label{fig3}
\end{center}
\end{figure}

As expected and detailed below, there are
continuous changes in the amount of ECM expansion over the life cycle. While we primarily examine the parental somatic cell layer and surrounding ECM, the 
developmental stage of the next generation is used for a precise definition of 
five key stages which we 
focused on in the comparative analyses (Fig. \ref{fig3}). Stage I: Freshly hatched young adults of 
equivalent circular radius (defined in \SI, \S 2.B) $R=106\pm 6\,\mu$m, containing immature gonidia. Stage II ($\sim 15$ h after hatching): 
Middle-aged adults with $R=221\pm 22\,\mu$m containing early embryos (4-8 cell stage).
Stage III ($\sim 21\ $h after hatching): Older middle-aged adults with $R=244\pm 15\,\mu$m containing embryos before inversion.
Stage IV ($\sim 36\ $h after hatching): Old adults of radius $R=422\pm 6\,\mu$m containing fully developed juveniles.
Stage S: Sexually developed adult females of radius $R=265\pm 29\,\mu$m bearing egg cells.
Since expression of PhII is induced by the sex-inducer protein 
\cite{RN5458,RN163,RN228},
in stages I to IV with vegetative phenotype, 
the sex-inducer protein was added 24 hours before microscopy to increase PhII expression; 
after such a short incubation with the sex inducer, the females still show an unchanged cleavage program and 
the vegetative phenotype. To obtain a changed cleavage program and a fully developed sexual phenotype (S), the females 
were sexually induced 72 hours before microscopy. 

\subsection{Pherophorin II is localized in the  compartment borders of individual cells}
\label{sec:phii_walls}

\begin{figure*}[t]
\begin{center}
\includegraphics[width=1.57\columnwidth]{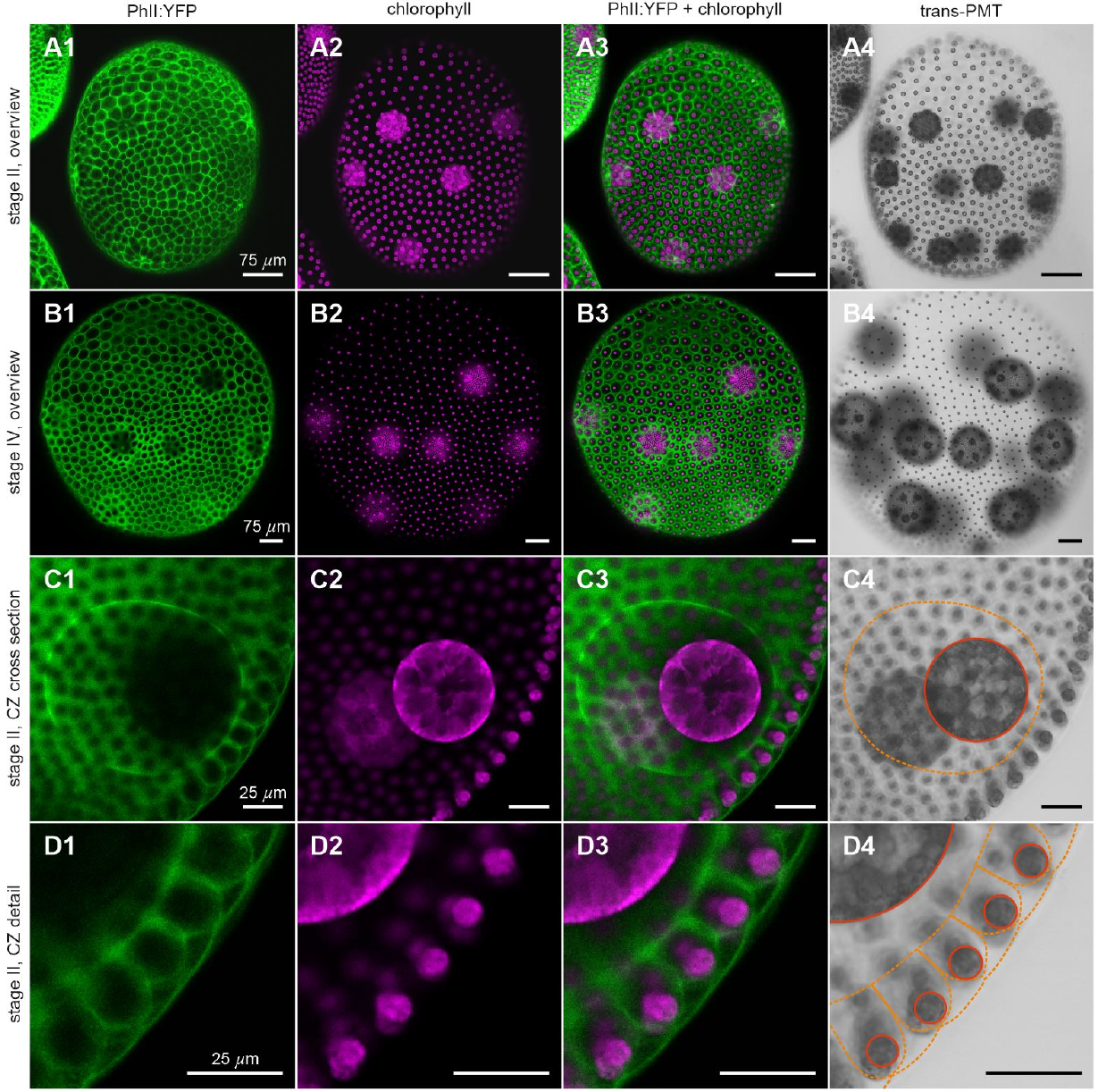}
\caption{Localization of PhII:YFP in whole, middle-aged (early stage II) and old adults (stage IV) in top view and magnified cross section through CZ. 
Sexually induced transformants expressing the \textit{phII}:\textit{yfp} gene under the control of the 
endogenous \textit{phII} promoter were analyzed in vivo for the localization of the PHII:YFP fusion protein. 
PhII:YFP is located in the CZ3 of both somatic cells and gonidia as well as in the BZ.
Top views of whole organism: (A) stage II immediately before the first cleavage of the gonidia inside; 
and (B) stage IV) before hatching of the fully developed juveniles. (C) Cross 
section through CZ in early stage II. (D4) Magnified view of the outer most ECM region. 
Column 1: YFP fluorescence of the PhII:YFP protein (green). 
Column 2: Chlorophyll fluorescence (magenta). Column 3: Overlay of YFP and chlorophyll fluorescence. 
Column 4: Transmission-PMT (trans-PMT). 
PhII:YFP-stained ECM boundaries are highlighted in orange and cell boundaries 
in red.}
\label{fig4}
\end{center}
\end{figure*}

\begin{figure*}[t]
\begin{center}
\includegraphics[width=1.75\columnwidth]{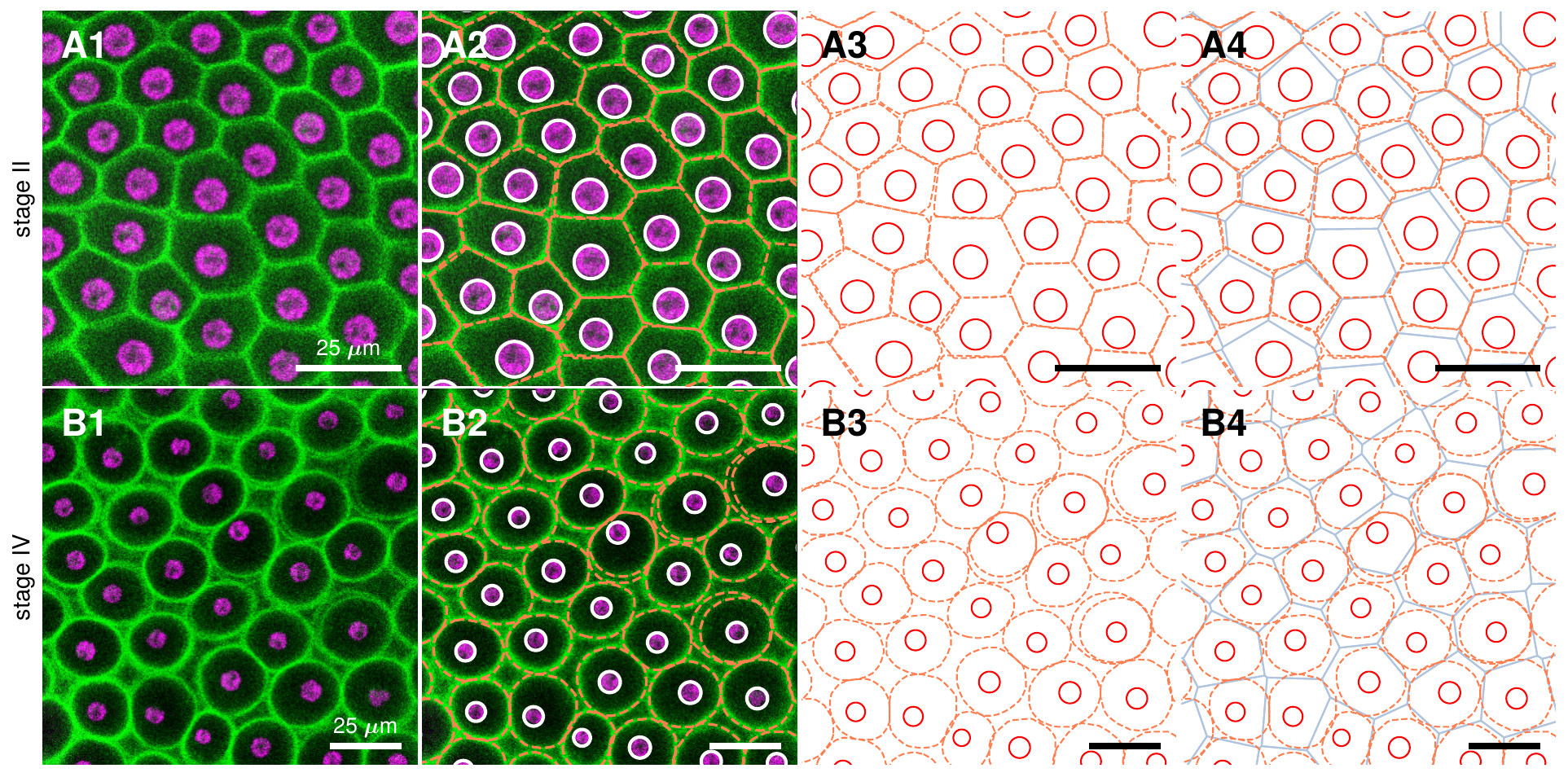}
\caption{As in Fig. \ref{fig4}, but a close-up of PhII:YFP localization in top view.
(A) Stage II. Magnified view of the somatic CZ3 compartments (1), identified by orange in 2 along with cell 
boundaries (white), shown together in 3 and in 4 with underlaid Voronoi tessellation (blue) and cell boundaries in red.
(B) In stage IV the CZ3 compartments become more bubble shaped and
individual CZ3 walls separate from the neighbors, leaving extracompartmental ECM space.
Double-walls also appear, highlighted in B3.
}
\label{fig5}
\end{center}
\end{figure*}

As expected, PhII:YFP is only found in the extracellular space within the ECM. It is detectable at all 
developmental stages after embryonic inversion, which marks the beginning of ECM biosynthesis
\cite{RN146,hallmann2000developmentally}, and 
in the ECM of organisms with the phenotypes of both vegetative and sexual development.
At a first glance, in a top view, PhII:YFP appears to form a polygonal pattern at the surface of each 
post-embryonic spheroid, with a single somatic cell near the center of each compartment (Fig. \ref{fig4}). 
PhII:YFP is also found in the ECM compartment boundaries (CZ3) of the gonidia, which are located below the 
somatic cell layer (Figs. \ref{fig1}B and \ref{fig4}).
These observations hold at all stages after embryonic inversion, even if the shape of the compartment boundaries varies.
On closer inspection (Fig. \ref{fig5}), it becomes apparent that: i)  the shapes of the somatic ECM compartment 
boundaries include a mix of hexagons, heptagons, pentagons and other polygons, as well as circles and ovals, 
ii) the angularity of the compartment boundaries changes in the course of expansion of the organism during development, 
i.e. the compartments become increasingly circular (less polygonal), and iii) each cell builds its own 
ECM compartment boundary. The observed localization of PhII is shown schematically in Fig. \ref{fig12}.

Especially in the early stages I and II of development, 
the fluorescent ECM compartment borders 
are predominantly pentagonal or hexagonal 
(Figs. \ref{fig4}A and \ref{fig5}A) with a rough ratio of 1:2 
(\SI, Fig. S4). By stage IV they become more rounded and 
the interstices we term ``extracompartmental ECM spaces'' between 
the compartments increase in size and number (Figs. 
\ref{fig4}B and \ref{fig5}B).
Since the compartment boundaries of adjacent cells 
are close to each other in early stages, 
the two compartments appear to be separated 
by a single wall. In later stages, when 
the boundaries are more circular, it becomes apparent that it is a double wall; thus, each somatic cell 
produces its own boundary (Figs. \ref{fig5}A,B).

As shown, PhII:YFP is a component of the CZ3 of both 
somatic cells and gonidia (Figs. \ref{fig1}B, \ref{fig4} and 
\ref{fig5}). It seems to be a firmly anchored building block
there, as the observed structure is highly fluorescent 
and yet sharply demarcated from other non-fluorescent 
adjacent ECM structures. If the PhII:YFP protein was 
prone to diffusion, one would expect a brightness 
gradient starting from the structures. There are no interruptions in the fluorescent labeling of boundaries associated 
with passages between neighboring compartments.
As the boundaries are so close together in 
early stages, only the thickness of the double wall 
can be determined and then halved; we estimate stage II single wall thickness as $1.6\pm 0.4\,\mu$m
and the similar value $1.9\pm 0.5\,\mu$m in stage IV. 

A comparison of the PhII:YFP-stained structures with 
descriptions of ECM structures from electron microscopy 
shows that the  PhII:YFP location corresponds exactly 
to that of the ECM structure CZ3 \cite{RN78}.
This applies to the entire period from the beginning of 
ECM biosynthesis after embryonic inversion 
to the maximally grown old adult. 

\begin{figure}[t]
\begin{center}
\includegraphics[width=\columnwidth]{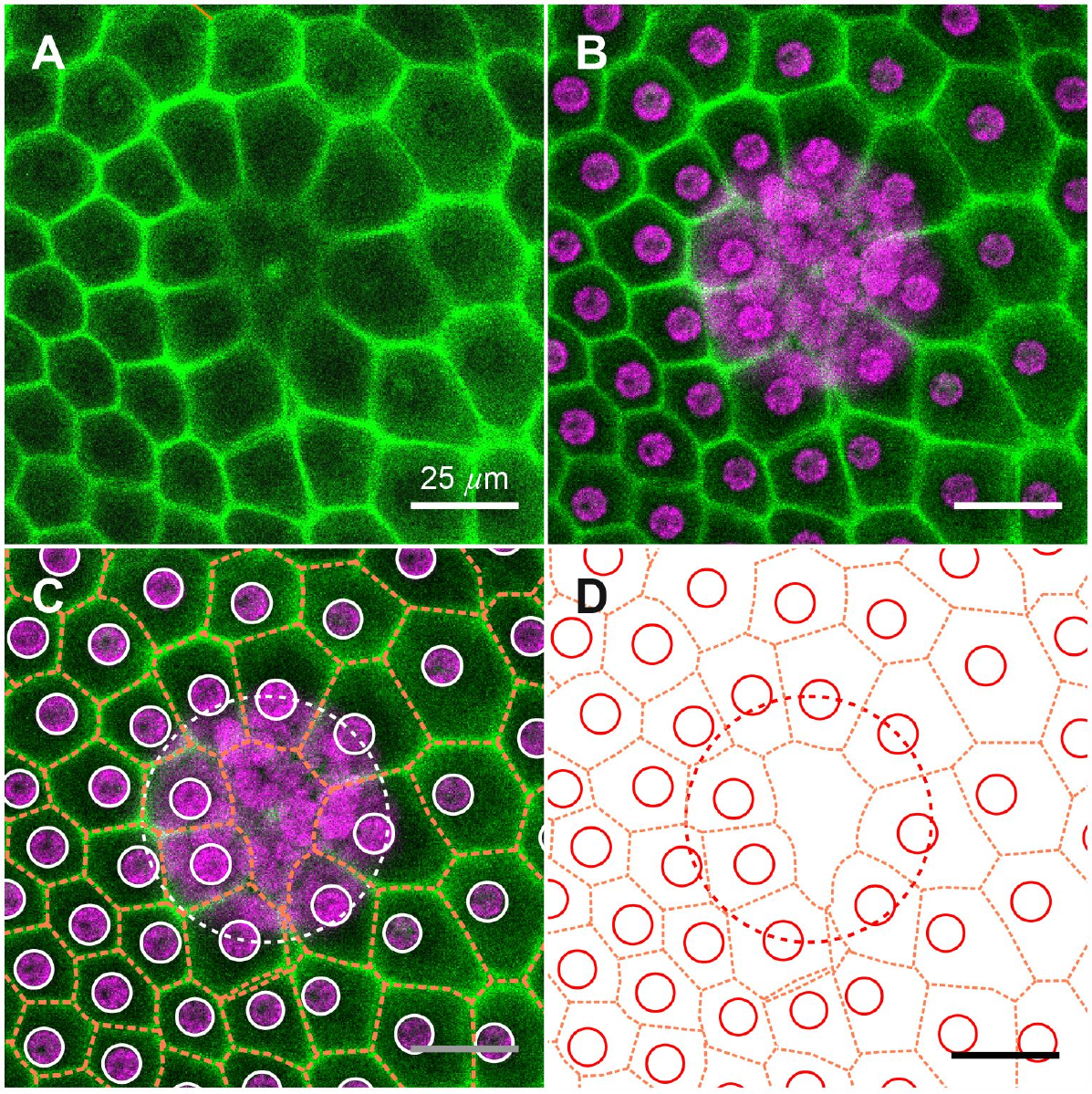}
\caption{Magnified top view of CZ3 above a reproductive 
cell in early stage II. (A) PhII:YFP fluorescence (green). (B) overlay 
with chlorophyll fluorescence (magenta). (C) overlay with highlighted CZ3 (orange) and cell 
boundaries (white). (D) as in (C) with only cell boundaries (red) and CZ3 (orange).}
\label{fig6}
\end{center}
\end{figure}

The relatively regular pattern of 
compartments (Fig. \ref{fig5}A) 
is disturbed by the gonidia (later embryos)
which, being far larger than somatic cells, are pushed under the somatic cell layer. The CZ3 compartments of the deeper gonidia nevertheless extend to the surface (Fig. \ref{fig6}). The surrounding somatic CZ3 compartments are elongated in the direction of the gonidial CZ3 protrusions to the surface (Fig. \ref{fig6}).
Since the PhII:YFP-stained CZ3 structure completely encloses each cell, these walls cannot be impermeable; even ECM proteins must pass through them. One indicator
for this is that the enormous growth of the spheroid during development requires the cell-free areas outside the ECM compartments of individual cells to increase immensely in volume. This applies in particular to the deep zone, but also to the areas between the compartments. 
All ECM material required for this increase in volume can only be produced and exported by the cells and must then pass through the ECM compartment borders of individual cells.

\subsection{Quantification of somatic CZ3 geometry}

The localization of PhII at the CZ3 
compartment boundaries allows us to carry out the 
first quantitative analyses of their geometry, both 
along the PA axis and through the life cycle 
stages.
A semi-automated image analysis pipeline 
(\SI, \S 2) reveals 
geometric features described in Table \ref{table:metrics} and Fig. \ref{fig7}.
A total of $29$ 
spheroids across five developmental stages 
were analyzed: 
$7$ in stage I, $5$ each in stages II, III, and IV 
of the asexual life cycle, and $7$ in the sexual 
life cycle (stage S).

The PA axis of \textit{V. carteri} spheroids 
corresponds to their swimming direction, and 
along this axis the distance between the somatic cells 
and the size of their eyespots decreases toward the 
posterior pole (Fig. \ref{fig1}A). Offspring (gonidia, 
embryos, and daughter spheroids) are mainly located in 
the posterior hemisphere.
We approximate the PA axis by a line passing through 
the center of the spheroid and normal to the best-fit 
line passing through manually identified juveniles in 
the anterior (Fig. \ref{fig7}A). 
This estimated PA axis is 
typically well-approximated by the elliptical 
major axis of the spheroid (\SI, Fig. S5).

The metrics presented in Fig. \ref{fig7} 
(see \SI, \S2)
are derived from the compartment shape outline or from moments of area.
The matrix ${\mathsf M}_2$ of second 
moments of area and the normalized matrix ${\mathsf \Sigma}={\mathsf M}_2/a_{\rm cz3}$, 
with the former defined as
\begin{equation}
    \label{eq:m2_cz3}
    {\mathsf M}_2 =\!\! \iint\limits_{\rm cz3} 
    ({\bm x} - {\bm x}_{\rm cz3})\!\otimes\!({\bm x} - {\bm x}_{\rm cz3}) d^2{\bm x},
\end{equation}
are interpretable as elastic strain tensors with respect to unit-aspect-ratio shapes. 
$\mathsf{\Sigma}$'s eigenvalues $\lambda_{\rm max}, \lambda_{\rm min}$ 
(whose square roots define the major, minor axis lengths) are principal stretches of this 
deformation, yielding the aspect ratio and other quantities defined in Table \ref{table:metrics}. 
Overall, we measure changes in the moments of area (\SI, \S 2.B) to quantify 
ECM geometry during growth;
the $0^\mathrm{th}$ gives the area increase, 
the $1^\mathrm{st}$ quantifies migration of  
compartment centroids with respect to cells, 
the $2^\mathrm{nd}$ gives the change in 
anisotropy (Table \ref{table:metrics}).
The sum of second moments reveals changes in crystallinity of the entire CZ3 
configuration, as shown in \S\ref{sec:tessellation}.

\ifPhIIpreprint\begin{table}[b]\else\begin{table}\fi
\centering
\ifPhIIpreprint\else\setlength{\tabcolsep}{3pt}\fi
\caption{Metric definitions.}
{
\ifPhIIpreprint\scriptsize\fi
\begin{tabular}{lccc}
metric & symbol & definition & units \\
\midrule
Somatic cell area & $a_{\rm cell}$ & - &\umsq \\
Somatic cell centroid & ${\bm x}_{\rm cell}$ & - & \um\\
CZ3 compartment area & $a_{\rm cz3}$ & -  & \umsq \\
CZ3 compartment centroid & ${\bm x}_{\rm cz3}$ & - & \um\\
CZ3 compartment perimeter & $\ell_{\rm cz3}$ & - & \um\\
CZ3 covariance matrix & ${\mathsf \Sigma}$ & \eqref{eq:m2_cz3} & \umsq \\
Aspect ratio & $\alpha$ & $\sqrt{\lambda_{\rm max}/\lambda_{\rm min}}$ & unitless\\
Circularity & $q$ & $\sqrt{4\pi a_{\rm cz3}}/\ell_{\rm cz3}$& unitless\\
Somatic cell offset vector & $\Delta {\bm x}$ & ${\bm x}_{\rm cell} - {\bm x}_{\rm cz3}$& \um\\
Somatic cell offset & $r$ & $\norm{\Delta {\bm x}}$& \um\\
Somatic cell offset (whitened) & $\Tilde{r}$ & $\sqrt{\Delta {\bm x}\cdot \Sigma^{-1}\Delta {\bm x}}$& unitless\\
Voronoi error & $e_V$ & vor $\cap$ cz3 / vor $\cup$ cz3 & unitless\\
\bottomrule
\end{tabular}
}
\label{table:metrics}
\ifPhIIpreprint\else\vspace{-5mm}\fi
\end{table}

\begin{figure}[t]
\begin{center}
\includegraphics[width=\columnwidth]{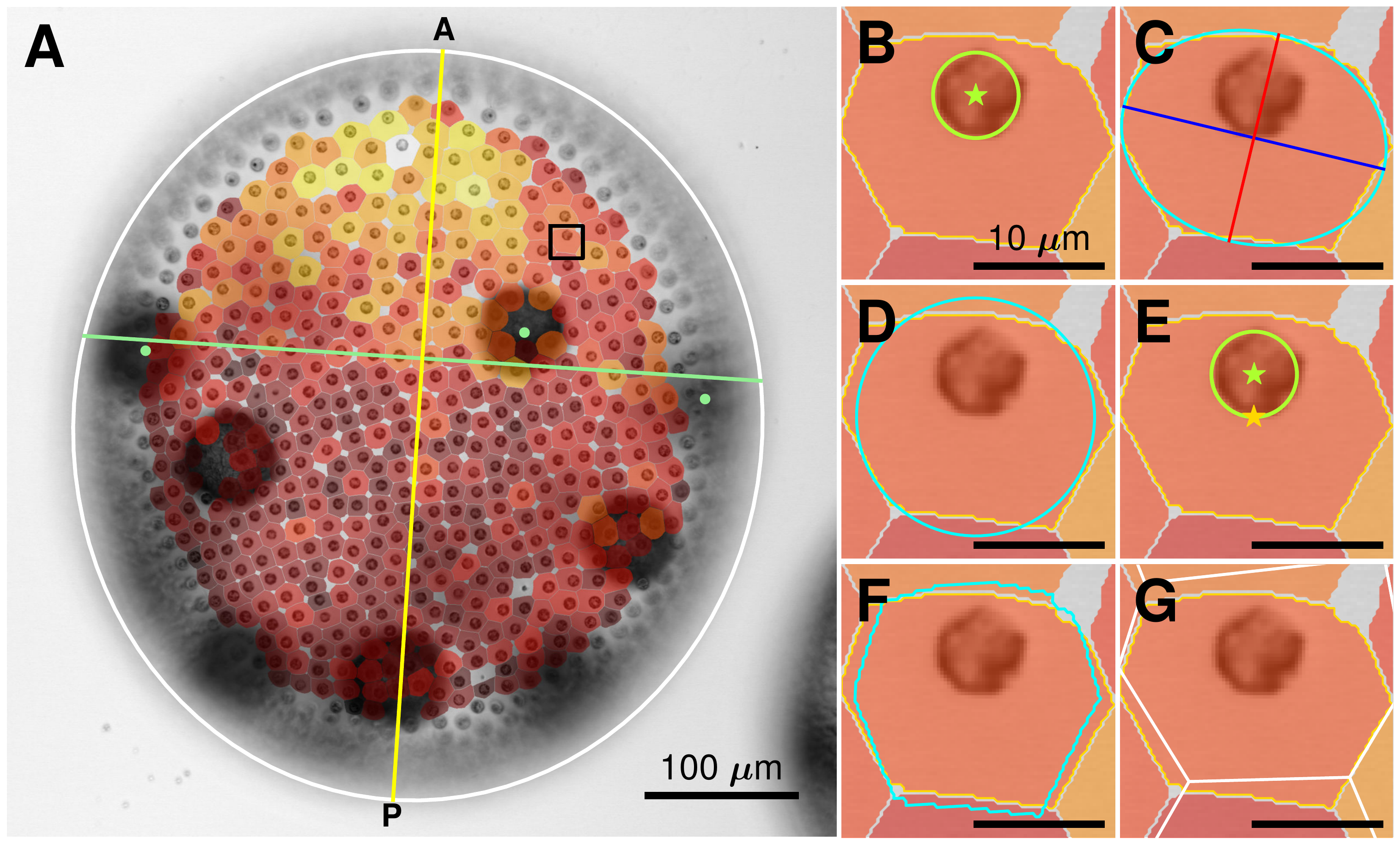}
\caption{Geometric features of cell/compartments pairs. (A) Trans-PMT image 
of stage III spheroid, with elliptical outline (white), estimated PA axis (yellow) 
that is orthogonal to line through gonidia (green dots).
Overlaid are segmentations of CZ3 compartments, colored dark to light by size.
(B-G) Schematics of geometric features computed from cell (green) and 
compartment (yellow) boundaries, as indicated: (B) $a_{\rm cell}, a_{\rm cz3}$, (C) aspect ratio $\alpha$ and 
corresponding ellipse (cyan), (D) deviation from a circle of the same area (cyan), (E) offset of cell center 
of mass (green star) from compartment center of mass (yellow star), (F) whitening transform of the 
compartment (cyan), and (G) Voronoi tessellation (white) 
error $e_V$.}
\label{fig7}
\end{center}
\end{figure}

\begin{figure}[htp] 
    \begin{minipage}[b]{0.48\textwidth}
    \includegraphics[width=\textwidth]{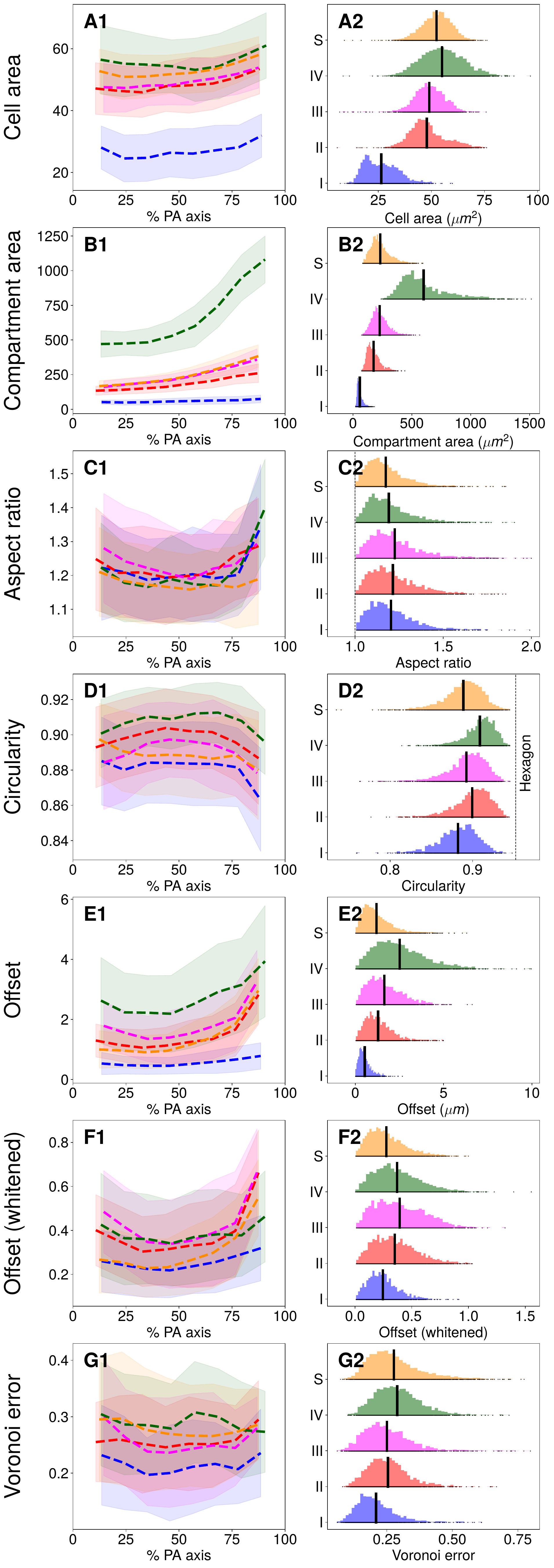}
    \end{minipage}%
    \hfill
\end{figure}

\begin{figure}[htp]
    \begin{minipage}[b]{0.48\textwidth}
\caption{PA axis and life cycle variation. (A-G) Column 1: Computed metrics binned in $8$ 
equally spaced segments along the PA axis.
Means are shown as dashed lines with per-bin standard deviation reported by shaded segments. 
Colors correspond to developmental stages defined in Fig. \ref{fig3}.
Column 2: Histograms of metrics by stage in $100$ equally spaced bins, by stage, with empirical 
means indicated by vertical black bars. 
Units are noted in parentheses, and otherwise, are dimensionless.}
        \label{fig8}
    \end{minipage}
\end{figure}

\subsection{CZ3 geometry along PA axis during the life cycle}

\subsubsection{\textbf{Anterior CZ3 compartments expand toward end of life cycle}}
\label{sec:cz3_ap}

Fig. \ref{fig8}A1-2 shows that the somatic cell area increases modestly, by $\sim 10\%$, along the PA axis at all stages. In contrast, the CZ3 compartment area grows substantially along this axis, with a minimum increase of $\sim 50\%$ in stage I and a maximum of $\sim 130\%$ in stages III and IV. 
Moreover, the slope increases after the equatorial region in all stages, an effect which is most pronounced in stage IV.
Cell and compartment areas also increase by life cycle stage as shown in Fig. \ref{fig8}, rows A-B. 
Somatic cell areas double from I-II, growing merely $\sim 10\%$ afterwards, whereas compartment areas expand primarily after III, with a $\sim150\%$ increase occurring from III-IV (\SI, Table S2). 
This surge in compartment area mirrors the spheroid's growth from III-IV, accounted an estimated $\sim 90\%$ to parental ECM volume changes, $\sim 10\%$ to that of growing juveniles, and minimally to that of somatic cells (Tables \ref{table:growth} and \SI, S1).

Fig. \ref{fig8}A2-G2 shows  
distributions of the metrics; apart from  
cell area, all exhibit positive skew and exponential 
tails which
suggest good fits with gamma-type distributions 
\cite{Day2022},
\begin{align}
    \label{eq:gamma}
    p_{\lambda, k}(x) &= \lambda^kx^{k-1}e^{-\lambda x} / \Gamma(k),
\end{align}
where $x$ is suitably standardized. 
This skew should be considered when making 
mean-based comparisons across life cycle stages. The 
long left tails of 
cell area reflect the persistence of small 
somatic cells throughout the life cycle, confirmed by inspection in the chlorophyll signal. Lastly, 
the cell size distribution primarily translates rightward in time, while the 
compartment size distribution simultaneously translates and stretches, indicating an increase in
polydispersity.

\setlength{\tabcolsep}{3pt}
\begin{table}[b]
\centering
{
\ifPhIIpreprint\scriptsize\else\tiny\fi
\begin{tabular}{lcccccc}
\toprule
\makecell{Stage \\ ~} & \makecell{Parent \\ radius} & \makecell{Offspring \\ radius} & \makecell{Somatic \\ cell radius} & \makecell{Parental ECM \\volume change} & \makecell{Parental ECM\\ growth rate} \\
& (\um) & (\um) & (\um) & (est., \mmcu) & (est., \mmcu/h) \\
\midrule
I   & $110 \pm 5.9$ & $16 \pm 2$ & $2.9 \pm 0.4$ & $\downarrow$ & $\downarrow$ \\
II  & $220 \pm 19$ & $29 \pm 2$ & $3.9 \pm 0.3$ & $0.039$ & $0.0026$ \\
III & $240 \pm 13$ & $30 \pm 4$ & $3.9 \pm 0.3$ & $0.015$ & $0.0025$ \\
IV  & $420 \pm 5.6$ & $79 \pm 6$ & $4.2 \pm 0.4$ & $0.23$ & $0.014$ \\
S   & $260 \pm 26$ & $15 \pm 1$ & $4.1 \pm 0.3$ & n/a & n/a \\
\bottomrule
\end{tabular}
}
\caption{Summary of estimated volumetric growth 
by stage.
Estimated ECM volume is volume of spheroid 
minus that estimated of juveniles and somatic cells, 
as explained in \SI, Table S1.
Values in final two columns represent changes with respect to preceding stage.}
\label{table:growth}
\end{table}

\subsubsection{\textbf{CZ3 compartments transition 
from tighter polygonal to looser elliptical packing}}
\label{sec:r2v}

While there is no apparent trend in the circularity 
of CZ3 compartments along the PA axis, the average 
circularity increases from stage I to IV. Since 
extracompartmental ECM space appears
as compartments increase in circularity
both effects correlate with enlargement of 
the spheroid. Fig. \ref{fig8}D2 shows that circularity 
increases in mean while decreasing in variance, 
suggestive of a relaxation process by which compartments 
of a particular aspect ratio but different polygonal 
initial configurations relax to a common elliptical shape with the same aspect ratio.
This transition is also apparent by the $\sim 39\%$
increase from stage I to IV in error with respect to the Voronoi 
tessellation (Fig. \ref{fig8}G2), 
whose partitions are always convex polygons.

\subsubsection{\textbf{CZ3 compartments enlarge anisotropically}}
\label{sec:cz3_aniso}

While the compartments become 
more circular as they expand, the aspect ratio 
is independent of stage 
and thus of organism size. The apparent increase at the 
ends of the posterior-anterior (PA) axis (U-shaped curves) is likely due to 
partially out-of-plane compartments appearing 
preferentially elongated.  
Fig. \ref{fig8}C shows that aspect ratio 
distributions are not only stable in mean, with 
less than $5\%$ variation, but also in 
skewness and variance; they are gamma-distributed throughout growth (\SI, Fig. S9).
Together, the 
invariance of aspect ratio and increasing 
compartment circularity during growth  
suggests a transition from tightly packed, polygonal 
compartments (where neighboring boundaries are closely 
aligned) to elliptical configurations in which 
neighboring boundaries are no longer in full contact.
We term this process \textit{acircular relaxation}.

To study how ECM is distributed around the somatic cells,
we quantified the cellular offset during
the life cycle. The absolute offset from the compartment 
center of mass (Fig. \ref{fig8}E1-2) increases 
from stages I to IV, and along the PA axis, 
indicating a strong correlation between larger 
compartment areas and cellular displacements (as we show in \S \ref{sec:corr}). 
Perhaps counterintuitively, the cellular offset vector shows no correlation with the primary elongation axis of the compartment and is uniformly distributed in $[0, \frac{\pi}{2}]$ throughout I-IV and S (\SI, Fig. S10).
In contrast to the absolute offset, the whitened offset (which takes into account compartment size and anisotropy) is nearly constant in mean after stage II 
(black vertical lines in Fig. \ref{fig8}F).
The support of the distribution does increase, 
albeit at a smaller rate than that of the cellular offset. 
Throughout this analysis of variation along the PA axis 
(Fig. \ref{fig8}), similarly sized spheroids in the 
asexual and sexual life cycle stages, 
bearing embryos or egg cells respectively, 
resemble each other in ECM geometry. 

\subsection{CZ3 geometry shows feature correlations}
\label{sec:corr}

The analysis above indicates compelling correlations between ECM growth and geometry. 
Here we analyze these in more detail with pooled data from all spheroids presented in Fig. \ref{fig8}.
Figure \ref{fig9}A shows an exponential increase in the compartment area $a_{\rm cz3}$ with cell area $a_{\rm cell}$ through stage III, saturating at stage IV.
This confirms that somatic cells primarily grow after hatching and before stage II, in contrast 
to compartments, which primarily grow after stage III.

As expected from the PA analysis, the aspect ratio (B) is decoupled from compartment size, 
while the circularity (C) increases.  This reinforces
the conclusion that as compartments expand they preserve their aspect ratio while decreasing in polygonality.
The cellular offset (D) reveals a 
power-law relationship with compartment area, which, in conjunction with the weak coupling between whitened offset and compartment size (E), further supports this conclusion in light of a scaling argument presented in Discussion \S5.

\begin{figure}[t]
\begin{center}
\includegraphics[width=\columnwidth]{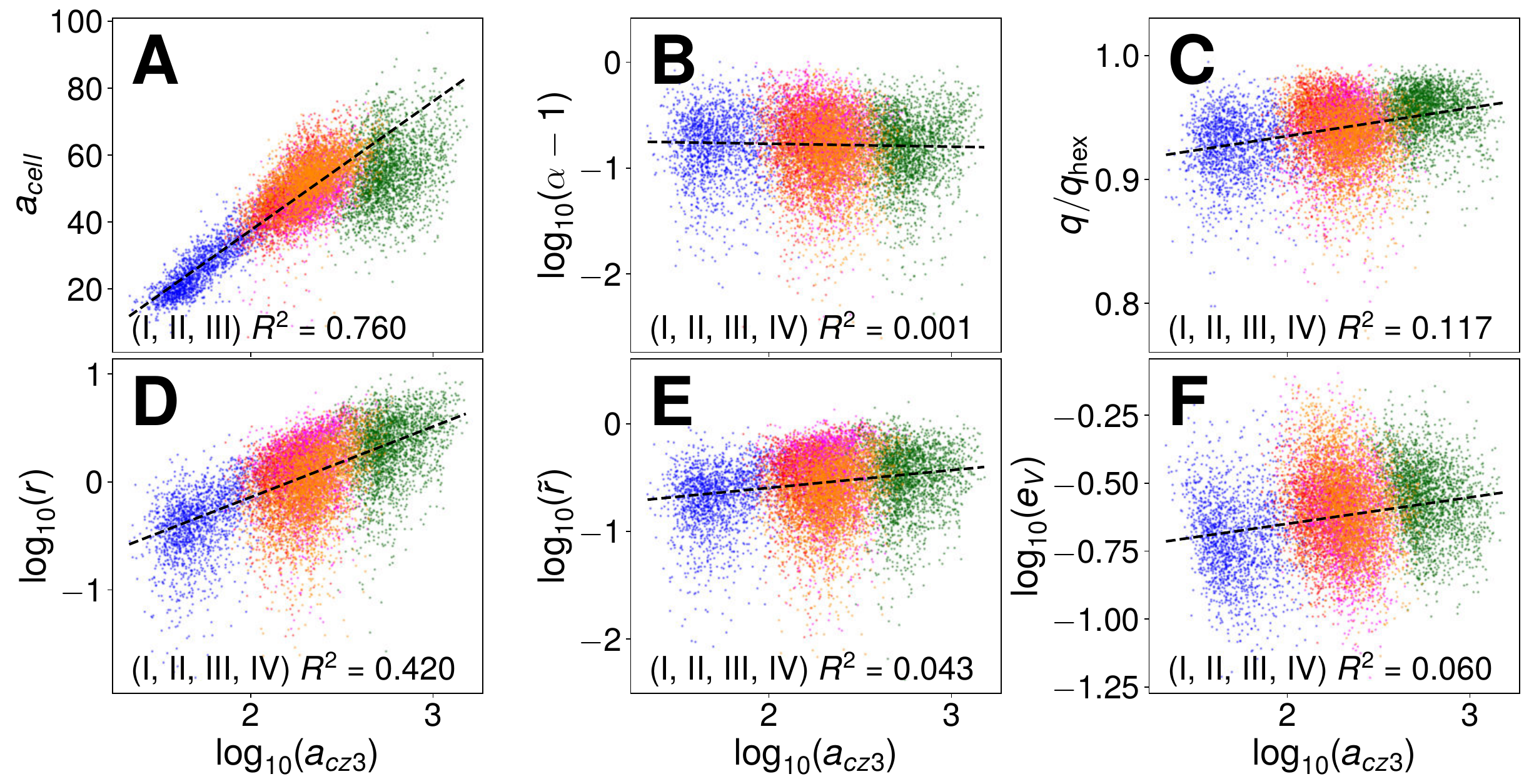}
\caption{Pair correlations of compartment features.  
At stages I-IV (blue, red, magenta, green) and S (orange), plots show correlations 
between compartment area ($a_{\text{cz3}}$) and metrics in Table \ref{table:metrics}
($q_{\rm hex}$ in (C) is the circularity of a regular hexagon).  
Coordinate transforms are chosen in either linear- or log-scale, with natural offsets, 
to produce most equally-sized contours across the distributions as measured via the 
kernel density estimate.
$R^2$ is the linear regression correlation coefficient for stages listed.}
\label{fig9}
\end{center}
\end{figure}

\subsection{Tessellation properties change during the life cycle}
\label{sec:tessellation}

The metrics in Fig. \ref{fig10} reveal clear trends by life cycle stage for the global geometry of each spheroid. 
Panel E shows the increasing circular 
radius, with most of the growth occurring between stages 
I-II and III-IV. II-III is separated by fewer hours and 
occurs during the first dark phase. Stage S is 
sorted in size close to stage III, supporting its resemblance with stages II and III in the PA analysis.

At fixed mean,
the shape parameter $k_{\rm gamma}$ of the gamma distribution \eqref{eq:gamma} is a measure of the entropy 
of the configuration,
with high $k$ indicating an increasingly crystalline, Gaussian-distributed configuration by the central limit theorem
\cite{durrett2019probability}. 
Fig. \ref{fig10}A confirms the stability of the aspect 
ratio distribution between stages, exhibiting 
values of $k_{\rm gamma}$ between $\sim2-3$, similar to ranges
previously reported for confluent tissues
\cite{atia2018geometric}. Simultaneously, panel B 
shows that $k_{\rm gamma}$ in the distribution of $a_{\rm cz3}$ is 
decreasing in stages I-IV, so the configuration (primarily the anterior hemisphere, \SI, Fig. S11) becomes increasingly disordered. The initial high 
values of $k_{\rm gamma}$ are consistent with the earlier 
observation that CZ3 compartments begin in a
tightly-packed configuration, and as $k$ quantifies  
regularity we infer that both tight packing and proximity 
to an equal-area lattice describes the initial 
configuration. The values of $k_{\rm gamma}$ between $2$ and $3$ in Stage IV
are close to those for the Voronoi tessellations 
\cite{Day2022}. This has implications for regime of 
validity of the Voronoi approximation, as discussed in 
\S \ref{sec:r2v}.

\begin{figure}[t]
\begin{center}
\includegraphics[width=0.9\columnwidth]{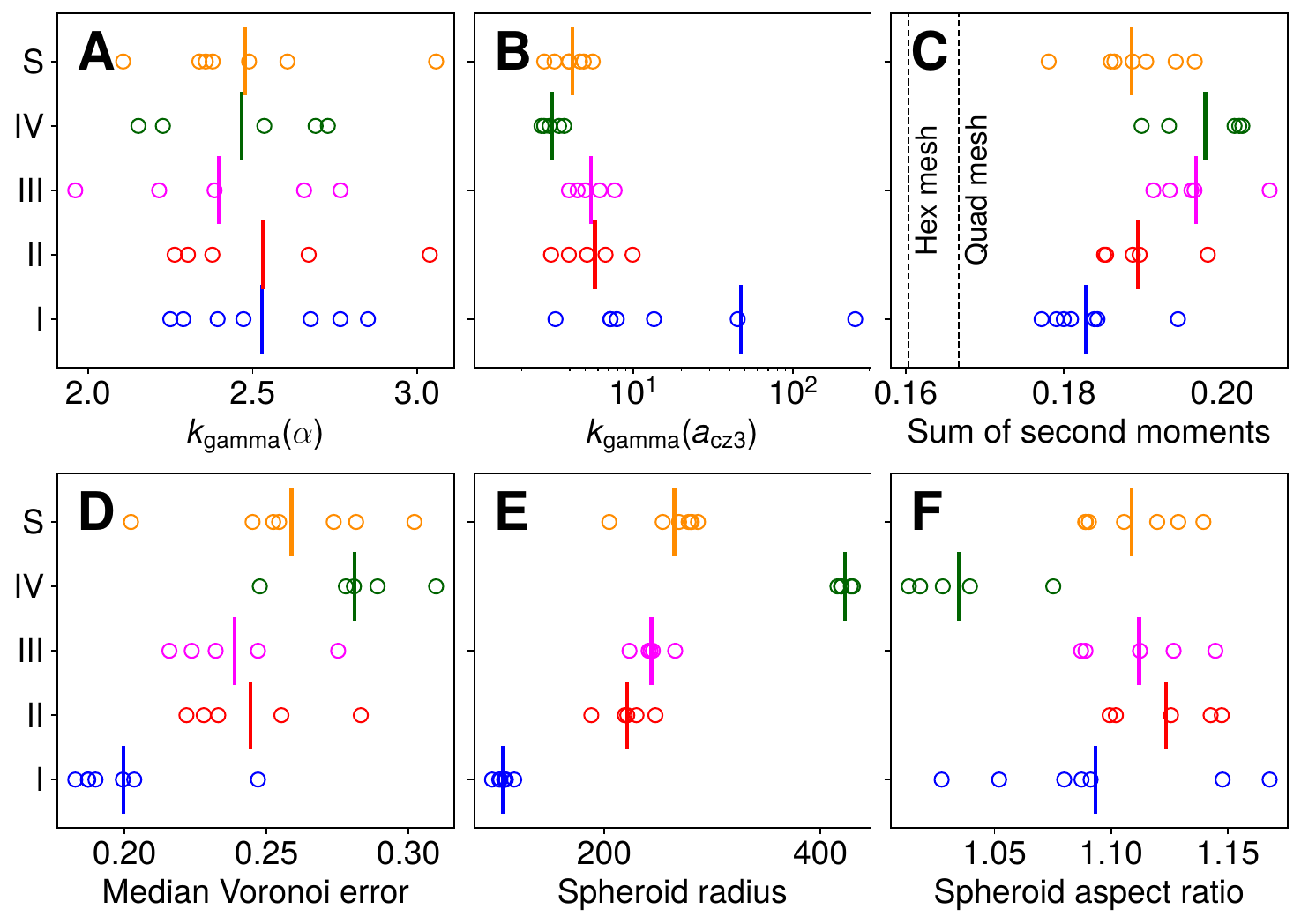}
\caption{\textbf{Global geometric properties of spheroids in stages I-IV, S.} (A) shape parameter 
$k$ of gamma distribution fit to $\alpha$, (B) the same for $a_{cz3}$, 
(C) sum of second moments \eqref{eq:sum_second_moments}, (D) median $e_V$ over the whole spheroid, 
(E) the circular radius of each 
spheroid (geometric mean of major and minor axes), and (F) the aspect ratio 
of each spheroid (ratio of major and minor axes).}
\label{fig10}
\end{center}
\end{figure}

The standardized sum of second moments, defined as
\begin{equation}
    \label{eq:sum_second_moments}
    n\sum_{i=1}^n\mathrm{Tr}(\mathsf{M}_{2}^{(i)}) / \left(\sum_{i=1}^n a_{\rm cz3}^{(i)}\right)^2,
\end{equation}
with $i=1\ldots n$ indexing CZ3 compartments per spheroid, 
is a space-partitioning cost that is minimized by equilateral hexagonal meshes (e.g. the surface of a honeycomb, see \SI, \S 2.C.2).
Although CZ3 
compartments do not tile space due to 
extracompartmental spaces, \eqref{eq:sum_second_moments} can nevertheless 
be computed for the covered area.
This energetic cost for each spheroid, encoding deviations from optimality arising from some form of perimeter excess, is displayed in panel C.
We find that somatic CZ3 configurations become \textit{decreasingly} optimal during expansion\textemdash 
an observation consistent with the underlying 
increasing trends in cell offset and area polydispersity  
observed in \S\ref{sec:cz3_ap} and \S\ref{sec:cz3_aniso}. 
The metrics in B-C thus show the 
counterintuitive result that the space partitioning formed by CZ3 compartments 
becomes increasingly disordered as the
spheroidal shape is maintained during the dramatic enlargement.

\subsection{Pherophorin II is also localized in the boundary zone}
Confocal cross sections reveal that PhII is also part of 
the boundary zone (BZ), the outermost ECM layer of the 
organism (Figs. \ref{fig1}B and \ref{fig4}C and D). The 
PhII:YFP-stained BZ extends as a thin 
$\sim 1.1\pm 0.4\,\mu$m arcing layer from 
the flagella exit points of one cell to those of all 
neighboring cells. This shape indicates that the 
outer surface of the spheroid has small 
indentations at the locations of somatic cells, 
where flagella penetrate the ECM. At these points, the 
BZ is connected to the CZ3 of the somatic cells below. 
Because the BZ is thin and not flat, it is not visible
in a top, cross-sectional view of a spheroid through 
the centers of the somatic ECM compartments 
(e.g. Fig. \ref{fig5}), and only partly visible when
the focal plane cuts through the BZ.
If the focal plane is placed on the deepest point of 
the indentations, only the areas at which the BZ is 
connected to CZ3 can be seen (Fig. \ref{fig11}). From the 
centers of these areas the two flagella emerge and the 
flagellar tunnels are seen as two black dots due to the 
lack of fluorescence there (Fig. \ref{fig11}B).
A closer look at the fine structure at the BZ-CZ3 
connection
site shows that fiber-like structures radiate from there 
to the BZ-CZ3 connection sites of neighboring
somatic cells. 

\begin{figure}[t]
\begin{center}
\includegraphics[width=\columnwidth]{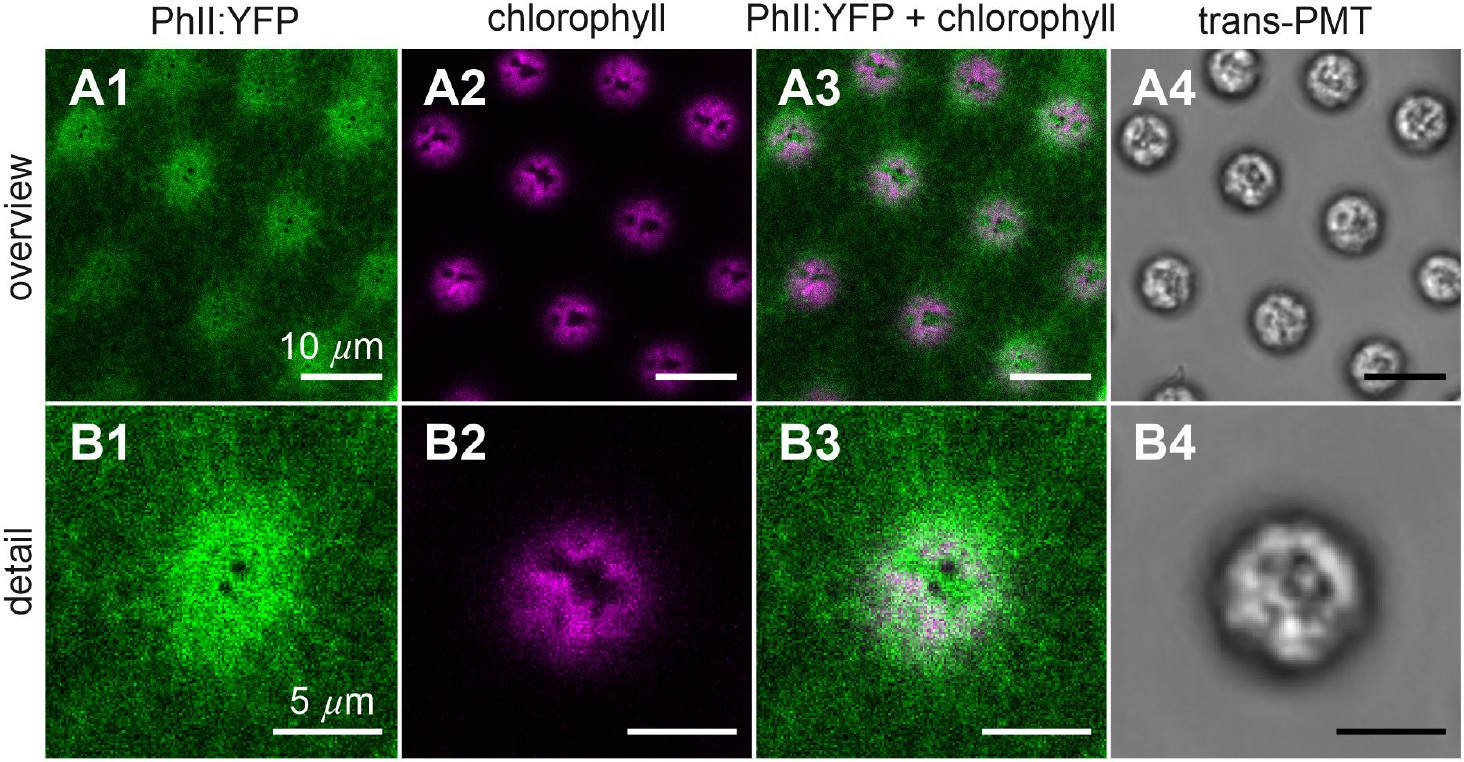} 
\caption{Magnified top view in regions
where the BZ is connected to 
CZ3 in early stage II.  Imaging as in Fig. \ref{fig4}. 
Fiber-like structures radiate from these regions.  Flagellar 
tunnels are seen in the centers of those areas as two dark dots.}
\label{fig11}
\end{center}
\end{figure}

\section*{Discussion and Conclusions}

\textbf{1. Holistic view of PhII localization.}
Synthesizing the results of preceding sections, we arrive
at the summary shown in Fig. \ref{fig12} of the 
identified locations of PhII.
It forms compartment boundaries (CZ3) not only 
around each somatic cell but additionally around each gonidium, 
and is also found in the outer border (BZ) of the spheroid;  CZ3 and BZ are 
connected where the flagella emerge. While each compartment boundary 
can be assigned to the cell it encloses, 
and is most likely 
synthesized solely by that cell, PhII in the BZ 
is evidently formed collectively by neighboring
cells. Since neither the compartment boundaries 
(CZ3) of the somatic cells are completely adjacent 
to the neighboring compartment boundaries, nor does 
the BZ rest directly on the compartment boundaries, 
extracompartmental ECM space remains between the 
CZ3 enclosures as well as between them and the BZ. 
The extracompartmental ECM space thus appears to be 
a net-like coherent space connected to CZ4.

\textbf{2. Relation to earlier ECM studies by electron microscopy.}
In earlier transmission electron microscopy images 
showing heavy metal-stained sections of the ECM, both 
the CZ3 and the BZ can be recognized as relatively dark 
structures, whereas the CZ2, CZ4 and the deep zone
are very bright \cite{RN78}. As the degree of darkness 
reflects the electron density and atomic mass 
variations in the sample, PhII evidently forms 
firmer wall-like structures in CZ3 and BZ, while  
CZ2, CZ4 and the deep zone have a very low density and 
are presumably of gel-like consistency. 
Using quick-freeze/deep-etch electron microscopy, 
it was shown that the fine structure of the ECM of 
volvocine algae such as \textit{Chlamydomonas} and 
\textit{Volvox} resembles a three-dimensional 
network \cite{RN222,RN158}.  While both the CZ3 and 
BZ are likely dense networks with a fine pore size
to the mesh, they must nevertheless allow the 
passage of small molecules and non-crosslinked 
ECM building blocks exported by cells, as evidenced by 
the growth during development of these
compartments, the extracompartmental space, and 
the deep zone. Cells must also be able to absorb 
nutrients from the outside, which must pass 
through both the BZ and CZ3. The BZ may be a denser
network than the CZ3, in order to prevent ECM building blocks from escaping into the environment.

\begin{figure}[t]
\begin{center}
\includegraphics[width=\columnwidth]{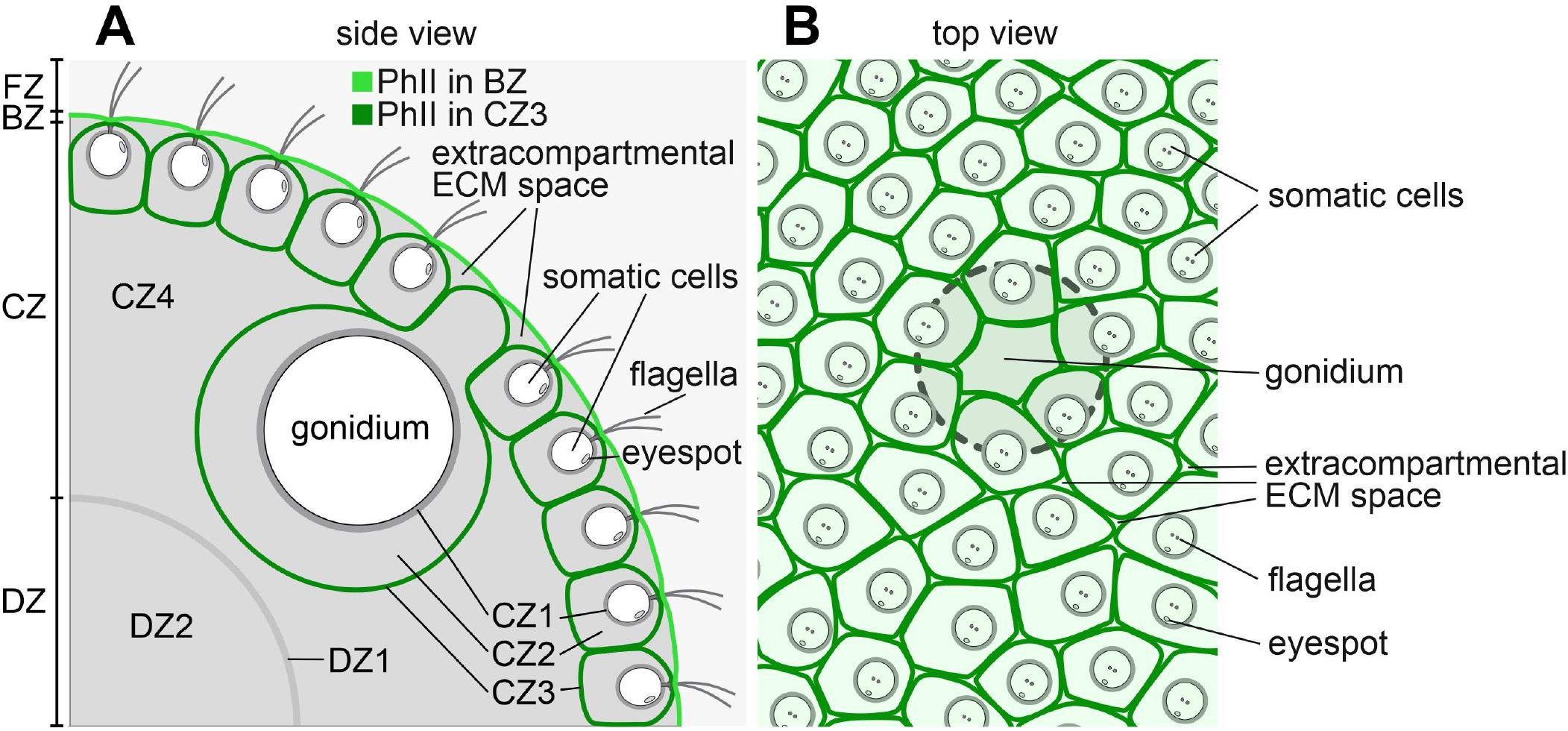}
\caption{Overview of PhII:YFP localization in the ECM in early stage II. (A) 
Schematic cross section, showing localization in CZ3 of both somatic cells 
and gonidia (dark green) and BZ (light green). (B) Schematic top view, looking 
through the boundary zone, showing
PhII:YFP in the CZ3 and the existence of 
extracompartmental ECM spaces. Position of a gonidium below somatic cell layer
is indicated (dashed); the CZ3 compartment of the deeper gonidium
extends to the surface where it is surrounded by ECM compartments with somatic cells.}
\label{fig12}
\end{center}
\end{figure}

\textbf{3. Mechanical implications of identified ECM structures.}
As revealed by the localization of PhII:YFP and prior
electron microscope studies, the BZ appears to 
form a dense ``skin'' on the outer 
surface of the spheroid to which the CZ3 compartments
are firmly attached 
at the flagella exit points. This 
point-like attachment allows the compartments to expand in all 
directions during growth. As the CZ3 compartments are 
densely packed and attached to the BZ, they would be 
expected to provide rigidity as a kind of 
``exoskeleton'' of the alga. A simple experiment
shows this feature: if many \textit{Volvox} 
are pressed together inside the suspending liquid medium 
and then released, they elastically repel 
each other. And while it is clear that the constant 
expansion of the compartments by incorporating further ECM 
components allows the ECM to enlarge considerably, 
the precise mechanism for transporting the ECM components to their destinations remains an open question.

\textbf{4. Evolution of the volvocine ECM and convergence of a monolayer epithelium-like architecture.}
The ECM of \textit{V. carteri} evolved from the (cellulose-free) cell wall of a \textit{Chlamydomonas}-like, 
unicellular ancestor \cite{RN896}. The cell wall of extant \textit{Chlamydomonas} species consists of an outer ``tripartite" layer with a highly regular, quasi-crystalline structure and an inner more amorphous layer. In few-celled volvocine genera with a low degree of developmental complexity, such as \textit{Pandorina}, the tripartite layer is partially split, so that its outer leaflet is continuous across the surface of the organism, while its inner leaflet still surrounds each individual cell body \cite{kirk1999evolution}. In larger, more complex volvocine algae (\textit{Eudorina} to \textit{Volvox}), the entire tripartite layer is continuous over the surface of the organism. In the genus \textit{Volvox}, the outer layer has developed even further and the tripartite layer has become part of the boundary layer, the BZ, while the inner layer has evolved species-specific (CZ3) compartments \cite{RN78}. The architecture of these CZ3 compartments shows certain parallels with the epidermis of most plant leaves \cite{esau1953plant} and even epithelia in animals \cite{dicko2017geometry}, all of which possess an (initially) closely packed, polygonal architecture. Looking at the plasma membranes in animal epithelia, cellulose-based cell walls in epidermal cells of land plants and cellulose-free CZ3 structures of somatic \textit{Volvox} cells as touching compartment boundaries, 
their packing geometry can be described using the same physical concepts \cite{lemke2021dynamic}. 
Interestingly, they also share the presence of an adjacent thin ECM layer, which represents a kind of boundary in all of them: the cuticle (secreted by plant epidermis cells), the basal lamina (secreted by animal epithelial cells) and the BZ (secreted by somatic \textit{Volvox} cells). The shared geometrical solution likely represents an example of convergent evolution driven by the common pressure to evolve a monolayer epithelium-like architecture with protective and control functions.

\textbf{5. Characteristics of somatic CZ3 geometry.} 
Examination of the stochastic geometry of the space partition formed by CZ3 walls,
along with somatic cell positions, reveals four key findings.
First, somatic cell growth occurs mainly between stages I-II, whereas CZ3 compartments 
grow mostly during III-IV (Fig. \ref{fig8}), indicating that ECM production, rather than 
somatic cell enlargement is the primary driver of visible compartment growth (Table 
\ref{table:growth}).
Their surface areas are well-approximated by gamma distributions (\eqref{eq:gamma}) 
with anterior/posterior hemispheres exhibiting different values of $k$ (\SI, Fig. S8).
Such area distributions arise in granular and cellular materials 
\cite{aste2008emergence,miklius2012analytical}; it is novel to observe them 
for intrinsic structures of an ECM, with non-stationarity in $k$ revealing 
A/P differentiation. 

Second, the aspect ratio distributions are remarkably invariant throughout the life cycle (Figs. 
\ref{fig8} and \ref{fig9}), and well-approximated by gamma distributions with stable $k$ 
(\SI, Fig. S9).
Maintaining a fixed $\alpha\approx 1.2$ requires that each compartment enlarges 
anisotropically, in strong contrast with the trend in aspect ratio of the overall spheroid, 
which is both lower and decreases from stages III-IV to less than $1.05$ (Fig. \ref{fig10}F). 
This lack of local-global coupling suggests that compartment anisotropy could be set by geometric 
constraints 
\cite{atia2018geometric} in the cellular configuration prior to stage I and
may explain how the organism maintains (and increases) its sphericity despite the strong 
non-uniformity in size and shape of its compartments.
Shape variability in the form of gamma-distributed aspect ratios arises in a large class of 
epithelial tissues and inert jammed matter \cite{atia2018geometric} which exhibit deviations from optimality in the sense of \eqref{eq:sum_second_moments}. The epithelium-like architecture formed by the CZ3 robustly falls into this class.

Third, the somatic cell offset increases steadily with compartment area (Fig. \ref{fig8}D) while the whitened offset remains relatively constant (Fig. \ref{fig8}E).
Both observations suggest a growth-induced deformation in which space is locally dilated (in a tangent plane containing the 
compartment) as $\R^2 \mapsto \rho \R^2, \rho \ge 1$, which indeed 
increases offsets while preserving whitened offsets (\SI, \S B.1).
Such transformations also naturally preserve the aspect ratio, consistent with earlier observations.
The cellular offset angle with respect to the principal stretch axis, on the other hand, 
is a priori unconstrained by these observations, and in fact we find that it is 
uniformly distributed in $[0, \pi/2]$ (\SI, Fig. S10).
This highlights a stochastic decoupling between cell positioning and compartment shape, much 
like that previously observed between compartment and spheroid principal stretches.
Again, this may be established before stage I and subsequently scaled by compartment growth\textemdash analogous to two points diverging 
on the surface of an expanding bubble.
Importantly, \textit{global} dilations of space, likening \textit{Volvox} itself to a bubble, 
cannot produce the increasing compartment area polydispersity we observe during the life cycle, 
while \textit{local} dilations as above, likening the CZ3 `epithelium' to a spherical raft 
of heterogeneously inflating bubbles, can.
In a continuum limit, such local dilations may be represented by conformal maps \cite{dai2022minimizing}.

Lastly, the CZ3 `bubbles' transition from a space \textit{partition} to a \textit{packing} whose fraction decreases 
from near-unity while ECM-filled extracompartmental spaces allow the compartments to relax from polygonal to elliptical shapes.
Such transitions are reminiscent of the hydration of foams \cite{phelan1995computation}, in which adding liquid 
(representing migration of ECM across the CZ3 walls, \S \ref{sec:phii_walls}) alleviates contact constraints.
Unlike classical foams, however, which relax rapidly to circular equilibrium shapes due to weak adhesion, 
inter-compartment adhesion is likely stronger due to crosslinking and the formation of insoluble networks.
Such surface tension-adhesion trade-offs can result in acircular equilibria with decreasing packing fraction 
\cite{kim2021embryonic}, unearthing one plausible explanation for the remarkable stability of aspect ratios as the 
compartments and extracompartmental spaces swell during growth (Fig. \ref{fig5}).
Notably, in contrast to typical assumptions of epithelial models \cite{bi2015density, kim2021embryonic}, the 
CZ3 `raft' is a self-assembled extracellular structure whose rheological properties 
including tension and adhesion are not under direct control by cells.
This highlights more broadly the need to probe the rheology of the ECM\textemdash perhaps using microrheological 
techniques akin to those applied in cytoskeletal studies \cite{microrheology}\textemdash to understand how stochastic 
local interactions give way to robust global structures.
Understanding the self-organization of crosslinking ECM components will shed light 
on the principles underlying the intricate geometry and stochasticity observed in multicellular extracellular matrices.

\matmethods{

\textbf{Strains and culture conditions.}
Female wild-type strains of \textit{Volvox carteri} f. \textit{nagariensis} were 
Eve10 and HK10. Eve10 is a descendant of HK10 and the male 69-1b, which originate from Japan. 
The strains have been described previously 
\cite{RN1592,RN145,RN137,RN68}. 
Strain HK10 has been used as a donor for the genomic library. As a recipient strain for 
transformation experiments a non-revertible nitrate reductase-deficient (\textit{nitA}$^-$) 
descendant of Eve10, strain TNit-1013 
\cite{RN5255},
was used. As the recipient strain is unable to use nitrate as a nitrogen source, it was 
grown in standard \textit{Volvox} medium 
\cite{RN144}
supplemented with 1 mM ammonium chloride (NH\textsubscript{4}Cl). Transformants with a 
complemented nitrate reductase gene were grown in standard \textit{Volvox} medium without 
ammonium chloride. Cultures were grown at 28°C in a cycle of 8 h dark/16 h cool fluorescent white light 
\cite{RN123}
at an average of $\sim$100 $\mu$mol photons m$^{-2}$ s$^{-1}$ photosynthetically 
active radiation in glass tubes with caps that allow for gas exchange or Fernbach flasks 
aerated with $\sim 50\,$cm$^3$/min of sterile air.

\textbf{Vector construction.}
The genomic library of \textit{V. carteri} strain HK10 in the replacement lambda phage vector $\lambda$EMBL3 \cite{RN150} described by \cite{RN19} has been used before to obtain a lambda phage, $\lambda 16/1$, with a 22 kb genomic fragment containing three copies of the \textit{phII} gene \cite{RN9}. A subcloned 8.3 kb \textit{Bam}HI-\textit{Eco}RI fragment of this lambda phage contains the middle copy, the \textit{phII} gene B, used here. The 8.3 kb fragment also includes the \textit{phII} promoter region, 5’UTR and 3’UTR and is in the pUC18 vector. An artificial \textit{Kpn}I side should be inserted directly upstream of the stop codon so that the cDNA of the \textit{yfp} can be inserted there. This was done by cutting out a 0.5 kb subfragment from a unique \textit{Mlu}I located 0.2 kb upstream of the stop codon to a unique \textit{Cla}I located 0.3 kb downstream of the stop codon from the 8.3 kb fragment, inserting the artificial \textit{Kpn}I with PCRs, and putting the \textit{Mlu}I-\textit{Cla}I subfragment back to the corresponding position. The primers 5’GTAACTAACGAATGTACGGC (upstream of \textit{Mlu}I) and 5’\textit{ATCGAT}TCA\underline{GGTACC}TGGCCCCGTGCGGTAGATG were used for the first PCR and the primers 5’\underline{GGTACC}\textbf{TGA}TTGCCGTAAGAGCAGTCATG and 5’TCTAGCCTCGTAACTGTTCG (downstream of \textit{Cla}I) for the second PCR (The \textit{Kpn}I side is underlined, the stop codon is shown in bold). One primer contains a \textit{Cla}I (italics) at its 5’end to facilitate cloning. PCR was also utilized to add \textit{Kpn}I sides to both ends of the \textit{yfp} cDNA. In addition, a 15 bp linker sequence, which codes for a flexible pentaglycine interpeptide bridge, should be inserted before the \textit{yfp} cDNA. The \textit{yfp} sequence was previously codon-adapted to \textit{C. reinhardtii} 
\cite{RN4929}
but also works well in \textit{V. carteri} \cite{RN5456}. Since this \textit{yfp} sequence was already provided with the linker sequence earlier \cite{RN5443}, the primers 5’\underline{GGTACC}\textit{GGCGGAGGCGGTGGC}ATGAGC and
5’\underline{GGTACC}CTTGTACAGCTCGTC and a corresponding template could be used (the \textit{Kpn}I side is underlined, the 15 bp linker is shown in italics). The resulting 0.7 kb PCR fragment was digested with \textit{Kpn}I and inserted into the artificially introduced \textit{Kpn}I side of the above pUC18 vector with the 8.3 kb fragment. All PCRs were carried out as previously described 
\cite{RN4265,RN4264,RN4263}
using a gradient PCR thermal cycler (Mastercycler Gradient; Eppendorf).
The final vector pPHII-YFP (Fig. \ref{fig2}) was checked by sequencing.

\textbf{Nuclear transformation of \textit{V. carteri} by particle bombardment.}
Stable nuclear transformation of \textit{V. carteri} strain TNit-1013 was performed as described earlier 
\cite{RN44} using a Biolistic PDS-1000/He (Bio-Rad) particle gun \cite{RN1098}.
Gold microprojectiles (1.0 $\mu$m dia., Bio-Rad, Hercules, CA, USA)
were coated according to earlier protocols \cite{RN4265,RN4264}.
Algae of the recipient strain where co-bombarded with the selection plasmid pVcNR15 
\cite{RN36}, carrying the \textit{V. carteri} nitrate reductase gene, and the non-selectable 
plasmid pPhII-YFP. Plasmid pVcNR15 is able to complement the nitrate reductase deficiency 
of the recipient strain and therefore allows for selection of transformants. For selection, 
the nitrogen source of the \textit{Volvox} medium was switched from ammonium to nitrate
and algae were then incubated under standard conditions in $9$ cm. diameter petri dishes 
filled with $\sim 35\,$ml liquid medium. Untransformed algae of the recipient strain die under 
these conditions due to nitrogen starvation. After incubation for at least six days, the
petri dishes were inspected for green and living transformants.

\textbf{Confocal laser scanning microscopy.}
For live cell imaging of transformed algae, cultures were grown under standard conditions 
and induced with 10 $\mu$l medium of sexually induced algae in a 10 ml glass tube. 
An LSM780 confocal laser scanning microscope 
was used with a $63\times$ LCI Plan-Neofluar objective and a $10\times$ Plan-Apochromat
(Carl Zeiss GmbH, Oberkochen, Germany). The pinhole diameter of the confocal was set to 1 Airy unit. 
Fluorescence of the PhII:YFP fusion protein was excited by an Ar\textsuperscript{+} 
laser at 514 nm and detected at 520-550 nm. The fluorescence of chlorophyll was 
detected at 650-700 nm. Fluorescence intensity was recorded in bidirectional scan mode for 
YFP and chlorophyll in two channels simultaneously. Transmission images were obtained in a third channel 
by using a transmission-photomultiplier tube detector (trans-PMT). Images were captured 
at 12 bits per pixel and analyzed using ZEN black 2.1 digital imaging software (ZEN 2011, Carl Zeiss GmbH). 
Image processing and analysis used Fiji (ImageJ 1.51w) \cite{RN5685}. 
To verify the signal as YFP fluorescence, the lambda scan function of ZEN was used
in which a spectrum of the emitted light is generated by a gallium arsenide phosphide 
QUASAR photomultiplier detector that produces simultaneous 18-channel readouts. 
Emission spectra between 486 and 637 nm were recorded for each pixel with a spectral resolution of $9\,$nm 
using a 458/514 beam splitter and 488-nm laser light for excitation. After data acquisition, 
spectral analysis was performed to allow separation of heavily overlapping emission signals.

\subsection*{Data, Materials, and Software Availability}
All data and code are available on Zenodo (\href{https://doi.org/10.5281/zenodo.14066435    }{DOI: 10.5281/zenodo.14066435})
\cite{Zenodo}.

}  

\showmatmethods{} 

\acknow{We are grateful to Kordula Puls and Diana Thomas-McEwen for technical assistance, and to Jane Chui and Kyriacos Leptos for inspiring conversations. Financial support for the work carried out in Bielefeld was provided by A.H.'s institutional funds. 
REG gratefully acknowledges the financial support of the John Templeton Foundation (\#62220).  This work was also supported in part by the Cambridge Trust (AS), and 
Wellcome Trust Investigator
Grants 207510/Z/17/Z (SSMHH and REG) and
307079/Z/23/Z (SKB, SSMHH and REG).}

\showacknow{} 
 
\bibliography{references}

\ifPhIIpreprint
    \ifArxiv
        \includepdf[pages={{},1,{},2,{},3,{},4,{},5,{},6,{},7,{},8,{},9,{},10,{},11,{},12,{},13,{},14,{},15,{},16,{},17,{},18,{},19,{},20,{},21,{}}]{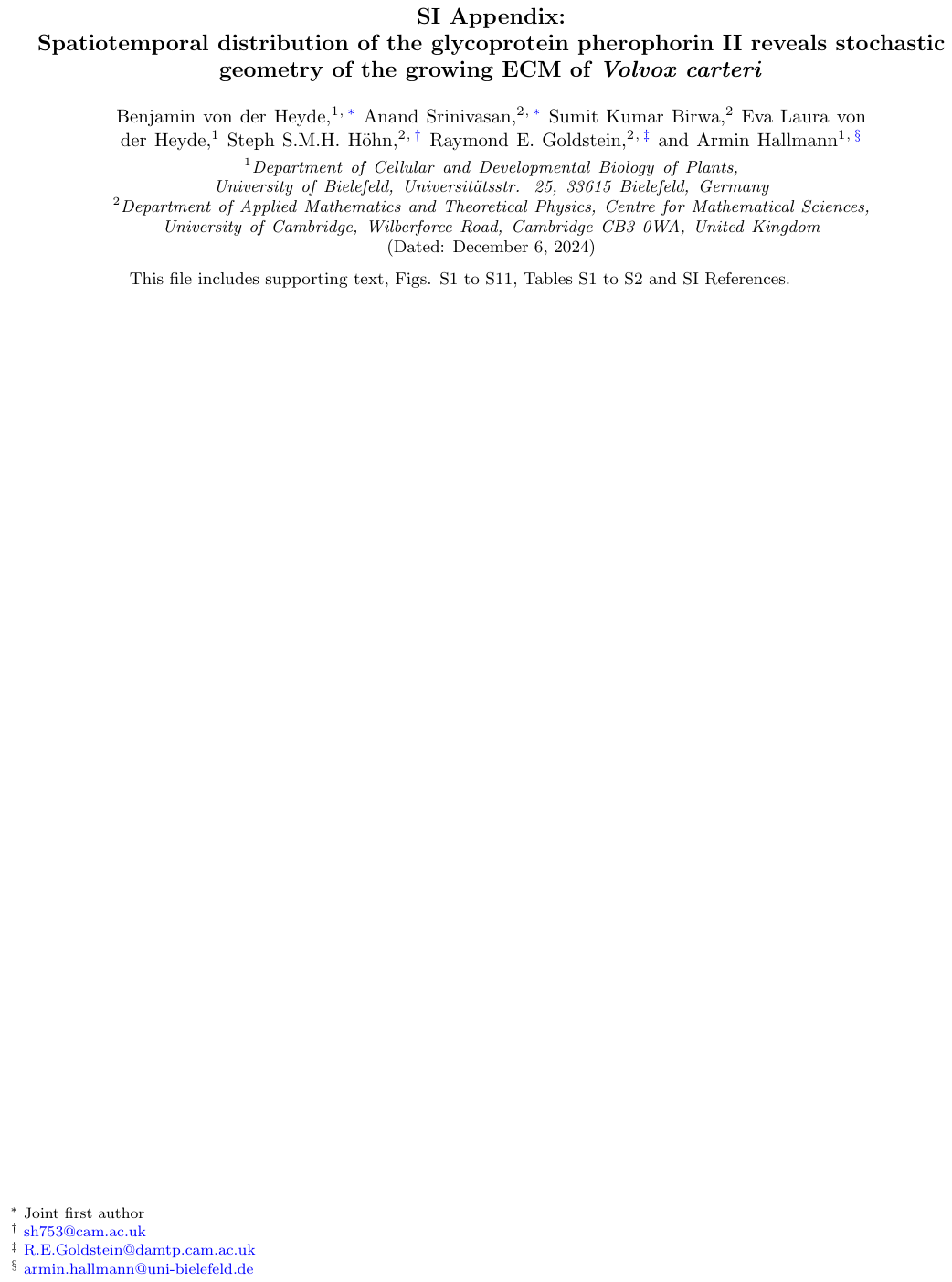}
    \else
        \includepdf[pages={{},-}]{PhII_SI_Ray.pdf}
    \fi
    \AtBeginShipoutNext{\AtBeginShipoutDiscard}
\fi
\end{document}